\documentclass[twocolumn]{aastex631}

\usepackage{amsmath}
\usepackage{bm}
\usepackage{booktabs}
\newcommand{\dif}{\mathrm{d}}

\newcommand{\Sg}{\Sigma}
\newcommand{\ag}{\alpha}

\newcommand{\kg}{\kappa}

\newcommand{\Om}{\Omega}

\newcommand{\pd}{\partial}

\newcommand{\hl}{{\bm {\hat l}}}
\newcommand{\hld}{{\bm {\hat l}}_{\rm d}}
\newcommand{\hlb}{{\bm {\hat l}}_{\rm b}}
\newcommand{\hs}{{\bm {\hat s}}}
\newcommand{\vlw}{{\bm {l}}_1}
\newcommand{\bG}{{\bm G}}
\newcommand{\btimes}{{\bm \times}}
\newcommand{\bcdot}{{\bm \cdot}}

\newcommand{\new}[1]{#1}
\newcommand{\newnew}[1]{#1}

\begin{document}

\title{Aligning Planet-Hosting Binaries via Dissipative Precession in Circumstellar Disks}

\correspondingauthor{Konstantin Gerbig}
\email{konstantin.gerbig@yale.edu}

\author[0000-0002-4836-1310]{Konstantin Gerbig}
\affiliation{Department of Astronomy, Yale University, New Haven, CT 06511, USA}

\author[0000-0002-7670-670X]{Malena Rice} 
\affiliation{Department of Astronomy, Yale University, New Haven, CT 06511, USA}

\author[0000-0002-9849-5886]{J.J. Zanazzi}
\altaffiliation{51 Pegasi b Fellow}
\affiliation{Astronomy Department and Center for Integrative Planetary Science, University of California Berkeley, Berkeley, CA 94720, USA}

\author[0000-0003-0046-2494]{Sam Christian}
\affiliation{Department of Physics and Kavli Institute for Astrophysics and Space Research, Massachusetts Institute of Technology, Cambridge, MA 02139, USA}

\author[0000-0001-7246-5438]{Andrew Vanderburg}
\affiliation{Department of Physics and Kavli Institute for Astrophysics and Space Research, Massachusetts Institute of Technology, Cambridge, MA 02139, USA}

\begin{abstract}

Recent observations have demonstrated that some subset of even moderately wide-separation planet-hosting binaries are preferentially configured such that planetary and binary orbits appear to lie within the same plane. In this work, we explore dissipation during the protoplanetary disk phase, induced by disk warping as the system is forced into nodal recession by an inclined binary companion as a possible avenue of achieving orbit-orbit alignment. We analytically model the coupled evolution of the disk angular momentum vector and stellar spin vector under the influence of a distant binary companion. We find that a population of systems with random initial orientations can appear detectably more aligned after undergoing dissipative precession, and that this process can simultaneously produce an obliquity distribution that is consistent with observations. \newnew{While dissipative precession proceeds efficiently in close binaries, favorable system properties (e.g., $r_{out} \gtrsim 100$ AU, $\alpha \gtrsim 0.05$, and/or $M_b/M_{*} \gtrsim 1$) are required to reproduce observed alignment trends at wider binary separations $a_\mathrm{b} \gtrsim450$ AU.} \newnew{Our framework further} predicts that circum-primary planets in systems with high stellar mass ratios should be preferentially less aligned than planets in equal-mass stellar binary systems. We discover \newnew{tentative evidence for this} trend in \textit{Gaia} DR3 and TESS data. \newnew{Our findings suggest} that dissipative precession \newnew{may play a significant role in} sculpting orbital configurations in \new{a sub-set of moderately-wide} planet-hosting binaries, \newnew{but is likely not solely responsible for their observed population-level alignment}. 
\end{abstract}


\section{Introduction}
\label{sec:intro}

Many host stars of known exoplanets also possess a moderate- to wide-separation stellar binary companion. The sky-projected orbital orientations within such s-type binary systems are diverse, ranging from fully aligned  \citep{Rice2023Qatar6, Rice2024} to polar configurations with high stellar obliquities \citep[e.g.][]{Winn2010Hatp11b, Sanchis2013, Esposito2014, Rubenzahl2021, Albrecht2021, Hjorth2021}. The \textit{Gaia} DR3 catalogue \citep{brown2022gaiadr3} allows stellar orbit fitting of planet-hosting binaries to an unprecedented precision, and has thus opened the pathway for population-level studies of their orbital geometries. A key finding is the trend of planetary orbits to preferentially appear aligned with the binary orbit
\citep{Christian2022align, Dupuy2022, Zhang2023dynamical, Lester2023, Rice2024, Christian2023inprep}: a statistically significant preference that holds even up to a binary separation of nearly $10^3$ AU. This trend can be interpreted as the combined outcome of the systems' birth orientations as resultant from star formation \citep{bate2010chaotic, Bate2018} and secular interactions that may occur `primordially', between the binary companion and circumstellar disk \citep{Batygin2012, Batygin2013, Lai2014, Spalding2014, XiangGruess2014, Zanazzi2017}, and/or post planet formation, between the binary companion and planet(s) \citep{Naoz2016, Zhang2018}. 

While star formation via disk fragmentation would naturally explain a preferential emergence of aligned systems, it is unlikely to be the dominant mode of binary formation outside of $\sim$ 500 AU \citep[see e.g.,][]{Bonnell1994, Tobin2018, Kuruwita2023, Offner2023}. On the other hand, core fragmentation of a (turbulent) molecular cloud core is able to produce a much wider range of binary orbital separations \citep[e.g.,][]{Padoan2002}, but results in largely randomly oriented initial configurations \citep{bate2010chaotic}.

We thus investigate the possibility of generating alignment --- or rather, damping misalignment --- during the protoplanetary disk epoch. This is an intuitive starting point, as gaseous protoplanetary disks possess viscosity and thus offer a straight-forward avenue to dissipate orbital energy. Our work is hence distinct from previous studies, such as \citet{Zhang2018}, which consider post-disk evolution of planetary embryos which may align (or anti-align) with the accompanying binary orbit through chaotic effects -- namely close encounters or direct collisions. Such processes work within $\sim 200$ AU only \citep{Zhang2018} and are thus likely insufficient to explain observed alignment trends at wider separations.

The paradigm under scrutiny instead considers a protoplanetary disk that is initially inclined with respect to the binary orbit. The binary companion drives the disk into differential nodal recession \citep{Batygin2012, Lai2014}, leading to warps. The resulting back-reaction torque evokes viscosity-dependent damping terms \citep{foucart2014evolution, Zanazzi2018torquedamp}. At the same time, the disk's coupling to the stellar spin vector has been identified as a potential source of obliquity excitation, which is expected if the recession rate of the stellar spin about the disk falls below the disk's recession rate about the binary. Therefore, the same mechanism jointly sculpts the spin-orbit distribution of exoplanets within binary star systems.

In this paper, we revisit and expand upon the considerations in \citet{Lai2014} and \citet{Zanazzi2018torquedamp}\new{, where warp amplitudes are assumed to remain small such that the disk is behaving \textit{almost} like a rigid body}. Unlike the individual systems probed in these previous works, our analysis specifically seeks to investigate the impact of \new{dissipative precession mitigated by viscosity} on a population of disks. For this purpose, we co-evolve the stellar spin and disk angular momentum vectors of a set of systems using a wide range of initial orientations and track the expected distribution of sky-projected orbital configuration angles.

In Section~\ref{sect:equations_of_motion} we present the relevant equations of motion. Section~\ref{sect:viscous_diss_and_disk_model} reviews viscous dissipation \new{in precessing disks} and highlights our disk model \new{within the context of the current observational consensus}.  Section~\ref{sect:obs_oriented_system} introduces the observer-oriented coordinate system, required to appropriately compare to observations of sky-projected orientations. Section~\ref{sect:results} contains the results of our analysis. Section~\ref{sect:obs_data} considers observational data \new{from \textit{Gaia} DR3 and the Transiting Exoplanet Survey Satellite (TESS), in the specific context of our model's predictions.} Section~\ref{sect:additional_alignment_pathways} outlines \new{additional effects which may sculpt orbital configurations in planet-hosting binaries,} and \new{our conclusions}  can be found in Section~\ref{sect:discussion}.

\section{Nodal Recession due to Binary Potential}
\label{sect:equations_of_motion}

We consider a primary star with mass $M_*$, radius $R_*$, and rotation rate $\Omega_*$ at the coordinate origin. The primary hosts a protoplanetary disk and co-orbits with a stellar binary companion with mass $M_\mathrm{b}$ at a semi-major axis $a_\mathrm{b}$. The binary orbit, characterized by the unit normal vector $\hat{\bm{l}}_\mathrm{b}$, torques the primary star's disk. This drives nodal recession of the disk and, if the primary star's spin axis is coupled to the disk, a corresponding recession of the stellar spin vector.

To model this process, we follow \citet{Zanazzi2018torquedamp} and track the evolution of the unit normal vector of the disk orbital angular momentum $\hat{\bm{l}}_\mathrm{d}$ and the unit vector of the primary star's spin axis $\hat{\bm{s}}$ via \citep[see also][]{Lai2014, Zanazzi2018}
\begin{align}
\label{eq:ld_evol}
\begin{split}
    \frac{\dif \hat{\bm{l}}_\mathrm{d}}{\dif t} = &- \tilde{\omega}_\mathrm{ds}( \hat{\bm{l}}_\mathrm{d} \cdot  \hat{\bm{s}})  \hat{\bm{s}} \times  \hat{\bm{l}}_\mathrm{d} \\ &- \tilde{\omega}_\mathrm{db}( \hat{\bm{l}}_\mathrm{d} \cdot  \hat{\bm{l}}_\mathrm{b})  \hat{\bm{l}}_\mathrm{b} \times  \hat{\bm{l}}_\mathrm{d} + \left(\frac{\dif \hat{\bm{l}}_\mathrm{d}}{\dif t}\right)_\mathrm{damp},
\end{split}\\
\label{eq:2_evol}
 \frac{\dif \hat{\bm{s}}}{\dif t} = &-\tilde{\omega}_\mathrm{sd} (\hat{\bm{s}}\cdot \hat{\bm{l}}_\mathrm{d}) \hat{\bm{l}}_\mathrm{d} \times \hat{\bm{s}} + \left(\frac{\dif \hat{\bm{s}} }{\dif t}\right)_\mathrm{damp}.
\end{align}
Here, $\tilde{\omega}_\mathrm{ds}, \tilde{\omega}_\mathrm{db}$ and $\tilde{\omega}_\mathrm{sd}$ are the recession rates of the disk around the star, the disk around the binary, and the star around the disk, respectively, and $({\dif \hat{\bm{l}}_\mathrm{d}}/{\dif t})_\mathrm{damp}$ and $({\dif \hat{\bm{s}}}/{\dif t})_\mathrm{damp}$ are damping terms for the disk orbit normal and the primary spin, respectively. The binary orbit is assumed be circular (eccentricity $e=0$), and $\hat{\bm{l}}_\mathrm{b}$ is assumed to remain constant in time.

We define the mutual inclinations between the set of unit vectors as
\begin{align}
    \theta_\mathrm{sd} &= \arccos(\hat{\bm{s}} \cdot \hat{\bm{l}}_\mathrm{d}), \\
    \theta_\mathrm{sb} & = \arccos(\hat{\bm{s}} \cdot \hat{\bm{l}}_\mathrm{b}), \\
    \theta_\mathrm{db} &= \arccos(\hat{\bm{l}}_\mathrm{d} \cdot \hat{\bm{l}}_\mathrm{b}).
\end{align}
These angles track the relative star-disk ($\theta_\mathrm{sd}$), star-binary ($\theta_\mathrm{sb}$), and disk-binary ($\theta_\mathrm{db}$) evolution as the system evolves.

\clearpage

\subsection{Recession rates}

The binary-driven recession rates are dependent on the system's stellar separation, orientation, and disk properties. Following \citet{Zanazzi2018}, we consider a rigidly rotating protoplanetary disk with surface density profile $\Sigma(r) \propto r^{-p}$, \new{for $r$ between the inner and outer truncation radii $r_\mathrm{in}$ and $r_\mathrm{out}$}. The disk is assumed to experience homologous mass loss, where the total disk mass evolves as \citep[see also][]{Lai2014, Zanazzi2018}
\begin{align}
\label{eq:mass_evolution}
    M_\mathrm{d} = \frac{M_\mathrm{d,0}}{1 + t/t_\mathrm{v}},
\end{align}
with initial disk mass $M_\mathrm{d,0}$ \new{and viscous timescale $t_\mathrm{v} = r_\mathrm{out}^2/\nu$. It is useful to express kinematic viscosity $\nu$ in terms of the Shakura-Sunyaev $\alpha$-parameter \citep[][]{Shakura1973} via $\nu = \alpha c_\mathrm{s}H$, where $c_\mathrm{s}$ is the sound-speed, $H = c_\mathrm{s}/\Omega$ is the disk pressure scale height, and $\Omega = \sqrt{GM_*/r^3}$ is the orbital frequency.}

For this model, the precession frequencies are given in \citet{Zanazzi2018} as
\begin{align}
    \tilde{\omega}_\mathrm{ds} &= \frac{3(5/2 - p)k_\mathrm{q}}{2(1+p)}\frac{R_*^2 \bar{\Omega}_*^2}{r_\mathrm{out}^{1-p}r_\mathrm{in}^{1+p}} \sqrt{\frac{GM_*}{r_\mathrm{out}^3}},\\ 
    \tilde{\omega}_\mathrm{db} &= \frac{3(5/2 - p)}{4(4-p)}\left(\frac{M_\mathrm{b}}{M_*}\right)\left(\frac{r_\mathrm{out}}{a_\mathrm{b}}\right)^3 \sqrt{\frac{GM_*}{r^3_\mathrm{out}}}, \label{eq:omdb} \\ 
    \tilde{\omega}_\mathrm{sd} &= \frac{3(2 - p)k_\mathrm{q}}{2(1+p)k_*}\left(\frac{M_\mathrm{d}}{M_*}\right) \frac{\bar{\Omega}_*\sqrt{GM_*R_*^3}}{r_\mathrm{out}^{2-p}r_\mathrm{in}^{1+p}},
\end{align}
where $k_*$ is the stellar spin normalization, $k_\mathrm{q}$ is the stellar quadrupole moment normalization, and $\bar{\Omega}_* = \Omega_*/\sqrt{GM_*/R_*^3}$ is the dimensionless rotation rate of the primary. \new{If the binary has nonzero eccentricity $e_{\rm b}$, $a_{\rm b}$ in Eq.~\eqref{eq:omdb} is replaced with $a_{\rm b,eff} = a_{\rm b}\sqrt{1-e_{\rm b}^2}$ \citep{Anderson+(2016)}. } Due to the mass loss of the disk, $ \tilde{\omega}_\mathrm{sd}$, initially the fastest precession rate, falls below $\tilde{\omega}_\mathrm{db}$ within a few viscous timescales. This \new{implies a secular resonance is crossed ($\tilde \omega_{\rm sd} \sim \tilde \omega_{\rm db}$) as the disk's mass depletes, which excites $\theta_\mathrm{sd}$ (and consequently stellar obliquities; see \citealt{Lai2014} for discussion). }

We can write the absolute value of the disk angular momentum as \citep{Zanazzi2018torquedamp}
\begin{align}
    L_\mathrm{d} = \frac{2 - p}{5/2 - p }M_\mathrm{d}\sqrt{GM_* r_\mathrm{out}}.
\end{align}
The fact that the binary orbital angular momentum $L_\mathrm{b}$ satisfies $L_\mathrm{b} \gg L_\mathrm{d}$ and also $L_\mathrm{b} \gg S = k_* M_* R_*^2 \Omega_*$ justifies treating the binary orbit as static.  

\section{Viscous dissipation}
\label{sect:viscous_diss_and_disk_model}

In this work, we are primarily interested in understanding the joint behavior of the spin-orbit angle $\theta_\mathrm{sd}$ and the orbit-orbit angle $\theta_\mathrm{db}$ (as well as the corresponding sky-projected angles that will be introduced in Section \ref{sect:obs_oriented_system}), the latter of which can only noticeably evolve if the system has a means of dissipating energy. Indeed, any loss of energy will move the system toward the lowest energy state -- that is, full alignment of the primary star's spin axis, disk orbit, and binary orbit, as seen in e.g. the Qatar-6 system \citep{Rice2023Qatar6, Rice2024}. Intriguingly, a polar configuration also has an absence of net-torque and thus constitutes an unstable equilibrium.

In this section, we \new{re-examine} the ability of a precessing protoplanetary disk to dissipate energy, quantifying the associated damping rates and dissipation timescales.

\subsection{\new{Small amplitude} disk warps}

\new{The formalism outlined in Sect.~\ref{sect:equations_of_motion} assumes that the disk remains effectively flat, in the sense that it precesses like a rigid body \citep[also see][]{Batygin2013, Lai2014, Spalding2015}. This rigid body approximation has to be enforced by effective communication between different disk annuli through either self-gravity or hydrodynamical effects.} 

\new{In this paper, we attribute this role to internal torques induced by small warps in the disk profile -- small amplitude bending waves \citep{Papaloizou1995, Lubow2000} -- which dissipate energy when resisting the external driving torque  \citep{Zanazzi2018torquedamp}, and ultimately result in the damping terms in Eqs.~\eqref{eq:ld_evol} and \eqref{eq:2_evol}. In order for our model to remain self-consistent, these warps have to remain small, which is the case as long as the bending wave crossing time is shorter than disk precession timescales. \citet{Zanazzi2018torquedamp} show that this is indeed an appropriate limit for the adopted framework. We thus proceed under the small warp approximation, which formally presupposes that the time-dependent component of the warp unit vector, i.e. the viscous twist, remains small: $|\bm{l}_1| \ll 1$. Note that this eliminates the possibility for disk breaking, which otherwise can result in a plethora of different dynamics \citep[e.g.,][]{Larwood1997, Lodato2010, Nixon2012, Dogan2015, Dogan2018, Smallwood2023}}.

\citet{Zanazzi2018torquedamp} provide semi-analytic expressions for the damping terms associated with \new{small amplitude} disk warps, i.e.
\begin{align}
\begin{split}
     \left(\frac{\dif \hat{\bm{l}}_\mathrm{d}}{\dif t}\right)_\mathrm{visc} & = \gamma_\mathrm{b}(\hat{\bm{l}}_\mathrm{d}\cdot \hat{\bm{l}}_\mathrm{b})^3 \hat{\bm{l}}_\mathrm{d} \times (\hat{\bm{l}}_\mathrm{b} \times \hat{\bm{l}}_\mathrm{d}) \\ &+ \gamma_\mathrm{s}(\hat{\bm{l}}_\mathrm{d}\cdot \hat{\bm{s}})^3 \hat{\bm{l}}_\mathrm{d} \times (\hat{\bm{s}} \times \hat{\bm{l}}_\mathrm{d}),
     \label{eq:dldt_visc}
\end{split}\\
     \left(\frac{\dif \hat{\bm{s}}}{\dif t}\right)_\mathrm{visc} & = - \frac{L_\mathrm{d}}{S}\gamma_\mathrm{s}(\hat{\bm{l}}_\mathrm{d}\cdot \hat{\bm{s}})^2 \hat{\bm{s}} \times (\hat{\bm{s}} \times \hat{\bm{l}}_\mathrm{d}),
     \label{eq:dsdt_visc}
\end{align}
where
\begin{align}
\begin{split}
\label{eq:damping_rate_binary_warps}
    \gamma_\mathrm{b} = & 1.26 \cdot 10^{-9} \left(\frac{\alpha}{10^{-2}}\right)\left(\frac{h_\mathrm{out}}{0.1}\right)^{-2} \left(\frac{M_\mathrm{b}}{1 M_\odot}\right)^2 \\ & \cdot\left(\frac{r_\mathrm{out}}{50 \mathrm{AU}}\right)^{9/2}  \left(\frac{a_\mathrm{b}}{300 \mathrm{AU}}\right)^{-6} \left(\frac{M_*}{1 M_\odot}\right)^{-3/2} \frac{2\pi}{\mathrm{yr}}
\end{split}
\end{align}
is the damping rate due to warps induced by the binary potential, and
\begin{align}
\begin{split}
\label{eq:damping_rate_oblate_star_warps}
    \gamma_\mathrm{s} = & 2.04 \cdot 10^{-10} \left(\frac{\alpha}{10^{-2}}\right)\left(\frac{h_\mathrm{in}}{0.1}\right)^{-2} \\ & \cdot \left(1358\frac{r_\mathrm{out}}{r_\mathrm{in}}\right)^{p-1}  \left(\frac{k_\mathrm{q}}{0.1}\right)^2 \left(\frac{\bar{\Omega}_*}{0.1}\right)^{4} \left(\frac{M_*}{1 M_\odot}\right)^{1/2} \\ & \cdot \left(\frac{R_*}{2 R_\odot}\right)^{4}   \left(\frac{r_\mathrm{in}}{8 R_\odot}\right)^{-4} \left(\frac{r_\mathrm{out}}{50 \mathrm{AU}}\right)^{-3/2} \frac{2\pi}{\mathrm{yr}}
\end{split}
\end{align}
is the damping rate due to warps induced by an oblate star. Notably, we ignore additional back-reaction terms, as the associated effect is dynamically negligible compared to that of $\gamma_\mathrm{b}$ and $\gamma_\mathrm{s}$ \citep[see][]{Zanazzi2018torquedamp}. \new{In the above scaling, we} write the disk aspect ratio as $h = H/r$, such that at the inner and outer truncation radius, $h_\mathrm{in} = H(r= r_\mathrm{in})/r_\mathrm{in}$ and $h_\mathrm{out} = H(r= r_\mathrm{out})/r_\mathrm{out}$, respectively. 

When plugging in fiducial model parameters (see the scaling relations in Eqs.~\eqref{eq:damping_rate_binary_warps} and \eqref{eq:damping_rate_oblate_star_warps}), \citet{Zanazzi2018torquedamp} find that, if $a_\mathrm{b} \gtrsim 200$ AU, viscous dissipation induced by disk warps is too weak to align the disk with the binary orbit plane within typical disk lifetimes. However, their modeled disks were relatively small and thus not necessarily representative of a typical (early stage) protoplanetary disk in a binary, especially given the high rate of large disks observed \citep[e.g.,][]{Najita2018, Boyden2020}. In addition, larger effective viscosities than $\alpha = 10^{-2}$ would further increase the damping rate. We discuss this prospect further in the next section. 

\subsection{Disk damping with high viscosity}
\label{sect:viscosity}

The viscosity of protoplanetary disks is relatively poorly constrained, ranging from $\alpha \sim 10^{-4}$ from observations of edge-on protoplanetary disks \citep{Villenave2020, Rosotti2023} or numerical calculations \citep[e.g.,][]{Flock2017} to $\alpha \sim 10^{-2}$, which well characterizes the viscous evolution of average protoplanetary disks \citep{Hartmann1998}. Indeed, \citet{Rafikov2017} argues that this range is indicative of an intrinsic viscosity distribution. Since these values were obtained by studying isolated disks which are likely not warped, it is conceivable that the associated $\alpha$-values are not necessarily appropriate for the disks in our study. In particular, disk warps can have resonant interactions with inertial waves, which result in a parametric instability \citep{Gammie2000, Ogilvie2013} that can increase viscosity and dissipation. If such an inertial wave resonance successfully operates in the precessing and warped disks in our study, the effective $\alpha$ values appearing in the damping rates in Eqs.\eqref{eq:damping_rate_binary_warps} and \eqref{eq:damping_rate_oblate_star_warps} may conceivably exceed the fiducial values in isolated disks by orders of magnitude. Therefore, in the following analysis, we will consider a \new{viscosity of $\alpha = 0.05$. For comparison, the fastest observed accretion rates are also associated with theoretically puzzling high $\alpha$-values up to 0.1 \citep[][\newnew{see also Sect.~\ref{sect:diskmodel_context}}]{Rafikov2017, Ansdell2018}.}

The bending wave regime only holds if $\alpha < h$ \citep{Papaloizou1983}. If, on the other hand, $\alpha > h$,  the disk lies in the diffusive regime. This subsection shows that the disk damping rates in both regimes are identical, so the same sets of evolutionary equations can be used regardless of the value of $\ag$ relative to $H/r$.

We start with the equations describing small warps in viscous disks, which satisfy $|\kg^2 - \Om^2|/\Om^2 \ll \ag^2 \ll 1$, where $\kg$ is the epicyclic frequency.  In this regime, the evolutionary equations for the warp may be written as \citep{Ogilvie1999,Dullemond2022}

\begin{align}
    \Sg r^2 \Om \frac{\pd \hl}{\pd t} &= \Sg {\bm T}_{\rm ext} + \frac{1}{r} \frac{\pd \bG}{\pd r}, \\
    \bG &\simeq \frac{\Sg H^2 r^3 \Om^2}{4} \left( \frac{1}{\ag} \frac{\pd \hl}{\pd r} + \frac{3}{2} \hl \btimes \frac{\pd \hl}{\pd r} \right).
\end{align}
Since $\ag \ll 1$, we will approximate
\begin{equation}
    \bG \simeq \bG_{\rm visc} = \frac{\Sg H^2 r^3 \Om^2}{4 \ag} \frac{\pd \hl}{\pd r}.
    \label{eq:Gvisc}
\end{equation}
The pressure term $\bG_{\rm press} = \frac{3\Sg H^2 r^3 \Om^2}{8} \hl \btimes \frac{\pd \hl}{\pd r}$ adds slightly to the warp profile, but is otherwise uninteresting dynamically.  Taking
\begin{equation}
    \hl(r,t) = \hld(t) + \vlw(r,t)
\end{equation}
with $|\vlw| \ll 1$, and adding external torques from the binary \citep[Eq.~1 of][]{Zanazzi2018torquedamp} and oblate star \citep[Eq.~3 of][]{Zanazzi2018torquedamp}, the same perturbation arguments can be repeated, substituting Equation~\eqref{eq:Gvisc} for Eq.~(17) of \cite{Zanazzi2018torquedamp}.  This leads to a viscous twist of
\begin{equation}
    \vlw = V_{\rm b} (\hlb \bcdot \hld)\hlb \btimes \hld + V_{\rm s}(\hs \bcdot \hld)\hs \btimes \hld.
    \label{eq:l1_visc}
\end{equation}
Here, $V_{\rm b}$ is given by Eq.~(31) of \cite{Zanazzi2018torquedamp}, while $V_{\rm s}$ is given by Eq.~(48) of \cite{Zanazzi2018torquedamp}.  Comparing our Equation~\eqref{eq:l1_visc} to Eq.~(59) of \cite{Zanazzi2018torquedamp}, we see that the viscous twist in the diffusive regime is identical to that in the bending wave regime.  Hence, Equations~\eqref{eq:dldt_visc} and~\eqref{eq:dsdt_visc} can still be used when $\ag \gtrsim H/r$, as long as $|\vlw| \ll 1$.

\begin{figure*}[t]
    \centering
    \includegraphics[width = \linewidth]{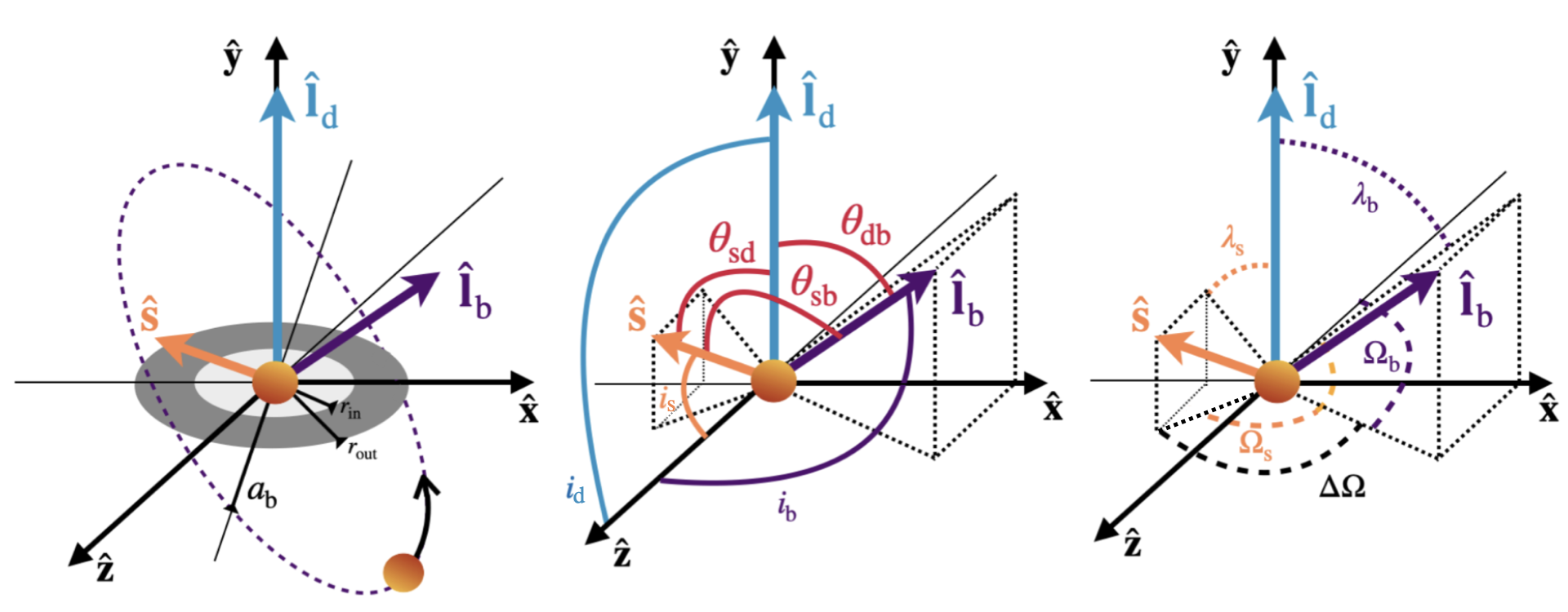}
    \caption{Schematic of the studied configuration. The left panel shows a protoplanetary disk with a binary companion. The three unit vectors $\hat{\bm{s}}, \hat{\bm{l}}_\mathrm{d}$ and $\hat{\bm{l}}_\mathrm{b}$ associated with stellar spin, disk orbital momentum, and binary orbital momentum respectively --- visible in all panels --- are drawn in the observer oriented coordinate system. The $z$-axis points towards the observer, \new{and fix the $y$-axis to lie parallel to $\hat {\bm {l}}_{\rm d}$, corresponding to $i_{\rm d} = 90^\circ$.  Longitudes (shown in dashed lines in the right panel) $\Omega_\mathrm{s}$ and $\Omega_\mathrm{b}$ define the angles between $-\hat{\bm z}$ with the projections of $\hat {\bm l}_{\rm b}$ and $\hat {\bm s}$ into the $\hat {\bm x}$-$\hat {\bm z}$ plane. Projected inclinations (shown in dotted lines in the right panel) $\lambda_\mathrm{s}$ and $\lambda_\mathrm{b}$ define the angles between $\hat {\bm l}_{\rm d}$ with the projections of $\hat {\bm l}_{\rm b}$ and $\hat {\bm s}$ onto the $\hat {\bm x}$-$\hat {\bm y}$ plane. Inclinations are displayed by solid arcs in the center panel.  The angles $\theta_{\rm db}$, $\theta_{\rm sd}$, and $\theta_{\rm sb}$ define the angles between the three unit vectors $\hat {\bm l}_{\rm d}$, $\hat {\bm l}_{\rm b}$, and $\hat {\bm s}$, while $i_{\rm d}=90^\circ$, $i_{\rm b}$, and $i_{\rm s}$ define the angles between the observer's direction $\hat {\bm z}$ with $\hat {\bm l}_{\rm d}$, $\hat {\bm l}_{\rm b}$, and $\hat {\bm s}$ respectively.}}
    \label{fig:angles}
\end{figure*}
\subsection{Suppression of Kozai-Lidov oscillations}
\label{sect:disk_kozai}

If the binary inclination relative to the disk is sufficiently large -- that is, if it falls within the range $39^\circ < \theta_\mathrm{db} < 141^\circ$ -- the disk can undergo Kozai-Lidov oscillations \citep{vonZeipel1910, Kozai1962, Lidov1962} where eccentricity and inclination of disk particles are exchanged in a manner such that $\sqrt{1-e^2}\cos i$ remains constant at the quadrupole level of approximation \citep[][]{Martin2014, Fu2015, Naoz2016, Lubow2017}. Using smoothed particle hydrodynamics calculations, \citet{Smallwood2023} argue that highly inclined disks are short-lived due to shocks that damp the disks' relative tilt within a few Kozai-Lidov timescales. 

On the other hand, in protoplanetary disks, gas pressure forces slightly eccentric orbits into apsidal precession, which suppresses Kozai-Lidov oscillations \citep{Wu2003}. \citet{Zanazzi2017} provide a criterion for the suppression of Kozai-Lidov instability due to pressure-induced apsidal precession, given by
\begin{align}
    h_\mathrm{out} \gtrsim  0.08 \cdot \left(\frac{5r_\mathrm{out}}{a_\mathrm{b}}\right)^{3/2} \left(\frac{M_\mathrm{b}}{M_*}\right)^{1/2}.
\end{align}
Disk self-gravity also drives apsidal precession of disk annuli, likewise working to suppress Kozai-Lidov oscillations \citep{Batygin2011}. Specifically, if the disk mass exceeds \citep{Zanazzi2017}
\begin{align}
\label{eq:KL_damping_mass_crit}
    M_\mathrm{d} \gtrsim 0.01 M_\mathrm{b} \cdot \left(\frac{5r_\mathrm{out}}{a_\mathrm{b}}\right)^{3},
\end{align}
Kozai-Lidov oscillations are suppressed. 

In this work, we are primarily concerned with the evolution of massive disks after the formation of the host star and the binary companion. Since we assume homologous mass loss via Eq.~\eqref{eq:mass_evolution}, we can rephrase Eq.~\eqref{eq:KL_damping_mass_crit} into a criterion on our time integration, i.e.
\begin{align}
    t \lesssim 200 t_\mathrm{v} \cdot \left(\frac{2M_\mathrm{d,0}}{M_\mathrm{b}}\right)  \left(\frac{5r_\mathrm{out}}{a_\mathrm{b}}\right)^{3},
    \label{eq:t_upper}
\end{align}
which, for $t_\mathrm{v} = 0.5$ Myr \new{(see Sect.~\ref{sect:diskmodel_context})}, lies well beyond the expected lifetime of disks. As such, we solve the evolution of systems where Eq.~\eqref{eq:KL_damping_mass_crit} holds such that Kozai-Lidov damping is suppressed by self-gravity. We forgo  including associated damping of relative inclinations in our evolutionary equations. We will revisit the Kozai-Lidov mechanism in Section~\ref{sect:kozai_planet} when discussing post-planet-formation systems without a disk that acts as suppressant.

\subsection{Scaling relations}
\label{sect:disk_scaling_relations}

In order to reduce the number of free parameters in our model, we employ scaling relations. Consider an equilibrium disk where the temperature scales as  \citep{Chiang1997, Dullemond2000, Wu2021}
\begin{align}
    T = 140 \mathrm{\ K}\cdot \left(\frac{M_*}{M_\odot}\right)^{-\frac{1}{7}} \left(\frac{L_*}{L_\odot}\right)^{\frac{2}{7}}\left(\frac{r}{1\mathrm{\ AU}}\right)^{-\frac{9}{20}}.
\end{align}
Using an isothermal sound speed of 
$c_\mathrm{s} =\sqrt{k_\mathrm{b}T/\mu m_\mathrm{p}}$,
the aspect ratio at the disk outer edge is
\begin{align}
\begin{split}
    h_\mathrm{out} = 0.08 \cdot \left(\frac{M_*}{M_\odot}\right)^{-\frac{4}{7}}\left(\frac{L_*}{L_\odot}\right)^{\frac{1}{7}}\left(\frac{r_\mathrm{out}}{100 \mathrm{AU}}\right)^{\frac{11}{40}},
\end{split}
\end{align}
assuming $\mu =2.33$ and provided the disk is locally isothermal, which is a reasonable assumption for passively heated disks \citep{Kratter2011}. The disk aspect ratio at the inner edge using the same scaling is $h_\mathrm{in} \sim 0.01$. 

Hotter stars are also more massive. Specifically, we consider a mass-luminosity relation $L_* \propto M_*^A$ with $A = 3/2$ following \citet{Chachan2023}, a choice motivated by MESA isochrone and stellar track models \citep{Dotter2016, Choi2016}. We thus have 
\begin{align}
\label{eq:hout_scaling_final}
    h_\mathrm{out} = 0.08 \cdot \left(\frac{M_*}{M_\odot}\right)^{-\frac{5}{14}} \left(\frac{r_\mathrm{out}}{100 \mathrm{AU}}\right)^{\frac{11}{40}}.
\end{align}
Note that, while more luminous stars host hotter disks, the corresponding increase in stellar mass is associated with an increased vertical gravity that leads to a net decrease in pressure scale height and consequently aspect ratio at a given radius. Specifically, we can re-write Eq.~\eqref{eq:damping_rate_binary_warps} as
\begin{align}
\label{eq:damping_rate_final}
\begin{split}
    \gamma_\mathrm{b} = & 4.45 \cdot 10^{-8} \left(\frac{\alpha}{10^{-2}}\right) \left(\frac{M_\mathrm{b}}{M_*}\right)^2 \left(\frac{M_*}{1 M_\odot}\right)^{\frac{17}{14}}  \\ & \cdot\left(\frac{r_\mathrm{out}}{100 \mathrm{AU}}\right)^{\frac{79}{20}}  \left(\frac{a_\mathrm{b}}{300 \mathrm{AU}}\right)^{-6}  \frac{2\pi}{\mathrm{yr}}.
\end{split}
\end{align}
For a constant binary mass ratio, having an overall massive and therefore hot system is advantageous for strong damping, because the damping rate has a stronger positive scaling with binary mass than its negative scaling with primary mass. 
The novel insight of Eq.~\eqref{eq:damping_rate_final} is that \new{inclination damping via} viscous dissipation operates more efficiently when the companion star is more massive than the host star. In other words, if the mechanism studied in this paper is the reason for observed alignment trends in planet-hosting binaries found in \citet{Christian2022align, Dupuy2022, Rice2024}, then circum-secondary planets (i.e. those orbiting the less massive binary component) are expected to have experienced more efficient alignment than circum-primary planets. We will test this hypothesis in Sect.~\ref{sect:unequal_massratio}.

\subsection{Disk model in the context of observed protoplanetary disks}
\label{sect:diskmodel_context}

\new{Our scaling for $h_{\rm out}$ in Eq.~\eqref{eq:hout_scaling_final} also sets the disk viscous time $t_{\rm v} \sim r^2/\nu$ evaluated at $r_\mathrm{out}$, i.e.,
\begin{align}
    t_{\rm v} \sim 0.50\ {\rm Myr} \cdot \left( \frac{0.05}{\ag} \right) \left( \frac{M_*}{M_\odot} \right)^{\frac{3}{14}} \left( \frac{r_{\rm out}}{100 \ {\rm AU}} \right)^{\frac{19}{20}} .
\end{align}
Because we take $r_{\rm out} = 100 \ {\rm AU}$ for all of our disks, and $t_{\rm v}$ scales weakly with $M_\star$, we approximate $t_{\rm v} \approx {\rm constant} = 0.5 \ {\rm Myr}$ for the rest of this work.}

The initial mass loss of a protoplanetary disk is primarily set by its viscous evolution, i.e. Eq.~\eqref{eq:mass_evolution}. On the other hand, its ultimate dispersal tends to be set by the photoevaporative mass loss rate, where far and extreme ultraviolet photons from the disk's host star produce hot electrons via photochemical reactions \citep[e.g.,][]{Ercolano2009, Owen2011}. Subsequent thermalization heats the gas to a degree sufficient to escape the star's potential well. Since X-ray luminosities increase with stellar mass, so does the photoevaporation timescale. Specifically, \citet{Komaki2021} find a mass loss rate of $\dot{M} \propto M_*^2$. We neglect this scaling in this paper and instead assume a mass-independent disk lifetime of $t_\mathrm{l} = 10^7$ yr. If considered, this dependence would influence the expectations for alignment distribution such that more massive systems would damp more quickly but for a shorter period of time, whereas less massive systems would damp inclinations more slowly but for a longer period of time. Combined with the scaling relationship $\gamma_\mathrm{b} \propto M_*^{17/14}$ between damping rate and total stellar mass in Eq.~\eqref{eq:damping_rate_final}, we expect this effect to be of negligible importance. We thus focus our study on the binary mass ratio --- on which the damping rate depends much more strongly --- rather than the total stellar mass.

\begin{table}[t]
    \centering
    \caption{Overview of constant parameters in our model. Other prescribed parameters are the masses of central star $M_*$ and binary companion $M_\mathrm{b}$, as well as their separation $a_\mathrm{b}$. The aspect ratio at the inner and outer truncation radii, $h_\mathrm{in}$ and $h_\mathrm{out}$, are calculated via Eq.~\eqref{eq:hout_scaling_final}.}
    \begin{tabular}{llr}
         Symbol & Meaning & Value \\ \midrule
         $R_\mathrm{s}$ & central star radius & $2 R_\odot$ \\
         $\overline{\Omega}_\mathrm{*}$ & normalized central star rotation rate & $0.1$ \\
         $M_\mathrm{d,0}$ & initial disk mass & $0.3 M_*$ \\
         $r_\mathrm{in}$ & inner truncation radius & $0.1$ AU \\
         $r_\mathrm{out}$ & outer truncation radius & $100$ AU \\
         $p$ & $\Sigma$ power law index & $1.0$ \\
         $k_\mathrm{q}$ & quadrupole coefficient & 0.1 \\
        $k_*$ & spin normalization coefficient & 0.2 \\ 
         $t_\mathrm{v}$ & viscous timescale & $0.5$ Myr \\
         $\alpha$ & disk viscosity & $0.05$ \\
         $t_\mathrm{l}$ & disk lifetime & $10$ Myr 
    \end{tabular}
    \label{tab:parameters}
\end{table}

\begin{figure}
    \centering
    \includegraphics[width=\linewidth]{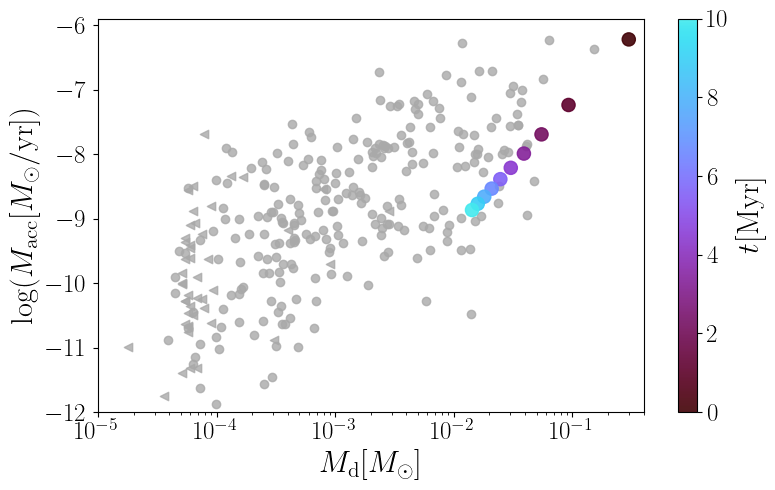}
    \caption{\new{Mass accretion rate and total disk mass for observed Class II disks in \citet{Manara2023}, with triangles corresponding to upper bounds in disk mass. Total disk mass is calculated multiplying the recorded mm continuum mass by a gas-to-dust mass ratio of 100. In color, we plot the trajectory of our disk model in this space following the viscous mass loss prescription in Eq.~\eqref{eq:mass_evolution} and the fiducial parameters in Tab.~\ref{tab:parameters}. For the model, accretion rate $M_\mathrm{acc} = |\dif M_\mathrm{d}/\dif t|$ is assumed to be due to viscous evolution only, as opposed to wind-driven accretion. This is not necessarily the case for the observed disks \citep[see][]{Manara2023}}.}
    \label{fig:diskmodel_comparison}
\end{figure}

Table~\ref{tab:parameters} shows an overview of constant parameters in our model, largely adopted from \citet{Zanazzi2018torquedamp}. Notable changes are the increased disk viscosity (see Sect.~\ref{sect:viscosity}) and a relatively large outer truncation radius of $r_\mathrm{out} = 100$ AU. \new{This value remains consistent with observations, which show a large variety of disk sizes from tens to hundreds of AU \citep{Najita2018}. The largest dust disks may extend up to $\sim 200$ AU \citep{Hendler2020}, and the gas disk is thought to be a factor of $\sim 2-3$ larger \citep{Ansdell2018}. \newnew{For such large disks, a viscosity parameter of $\alpha \sim 3\cdot 10^{-3}$ would, per Eq.~\eqref{eq:damping_rate_final}, yield the same damping rate as for our fiducial model with $\alpha = 0.05$.} The choice of $r_\mathrm{out} = 100 \mathrm{AU}$ \newnew{also} ensures that all models considered in this paper strictly adhere to $a_\mathrm{b}/r_\mathrm{out} \geq 4$, avoiding disk truncation by the binary orbit. Note that, if the binary's orbit were eccentric, $a_{\rm b,eff}/r_{\rm out} = a_{\rm b} \sqrt{1-e_{\rm b}^2}/r_{\rm out} \ge 4$ would avoid tidal truncation until $e_{\rm b} \gtrsim 0.7$, since the disk's tidal truncation radius $r_{\rm tide} \approx 0.35 (1-e_{\rm b})a_{\rm b} = 0.35[(1-e_{\rm b})/(1+e_{\rm b})]^{1/2} a_{\rm b,eff}$ \citep[e.g.][]{Miranda+2015}. } 

The initial stages of the \new{calculated} evolution aim to reflect the transition from Class I to Class II, hence the high initial disk mass of $M_\mathrm{d,0} = 0.3 M_*$ \citep[compare with e.g.][]{Andersen2019}, which is then rapidly reduced following the utilized prescription for homologous mass loss. \newnew{This disk mass renders the outer regions of the modeled disk (beyond $45$ AU) gravitationally unstable for $\lesssim 0.8 t_\mathrm{v} \sim 0.4$ Myr, possibly increasing turbulence $\alpha$ \citep[e.g.,][]{Gammie2001, Booth2019} and thus the damping rate $\gamma_\mathrm{b}$.} \new{We show the trajectory of our model in total disk mass $M_\mathrm{d}$ -- accretion rate $M_\mathrm{acc}$ space in Fig.~\ref{fig:diskmodel_comparison}. For comparison, we show observational data of Class II disks with corresponding measurements compiled by \citet{Manara2023} (their Fig.~7). Disk (gas) mass $M_\mathrm{d}$ is calculated from the dust mass traced by mm continuum emission assuming an interstellar medium-like solid-to-gas ratio of $1\%$.} 

\new{This approach relies on assumptions on uncertain dust opacities \citep[e.g.,][]{Wright1987, Andrews2005, Birnstiel2018, Liu2022, Miotello2023}, and it presupposes the mass distribution of solids to peak at mm sizes \citep{Williams2011, Manara2023}. If, however, radial drift, or aggregation via dust growth, and/or planet formation operate even moderately fast, some fraction of the total solid mass may be either lost or hidden in larger dust grains or planetesimals \citep[][]{Hughes2012, Gerbig2019}, leading to very different true solid-to-gas mass ratio and a potentially substantial under-estimation of total disk masses \citep[e.g.,][]{Birnstiel2012, Powell2019}. With this in mind, Fig.~\ref{fig:diskmodel_comparison} characterizes our model as a moderately massive and large, relatively slowly accreting disk that is well within the bounds of plausability established by the current observational and theoretical consensus.} 

\new{On first glance, the low accretion rates of the model relative to typical disks in the corresponding mass range appear to be inconsistent with the fairly high viscosity of $\alpha = 0.05$. Indeed, naively taking for example Eq.~(4) in \citet{Rafikov2017}, i.e. 
\begin{align}
    \label{eq:alphainferred}
    \alpha \sim \frac{M_\mathrm{acc}}{M_\mathrm{d}} \frac{\Omega r^2}{c_\mathrm{s}^2} 
\end{align}
evaluated at $r_\mathrm{out}$, one would infer an $\alpha$ that decreases in time from $\sim 0.08$ to $\sim 0.005$. This is because our model assumes a polynomial mass loss via Eq.~\eqref{eq:mass_evolution} as opposed to the exponential decay assumption that underlies Eq.~\eqref{eq:alphainferred} -- a reminder that subtleties in the assumed viscous evolution model may lead to appreciable differences when connecting accretion rates and viscosity.}

\newnew{To conclude this section,} in the construction of our fiducial model, we have deliberately selected parameters \newnew{that lead to favorable conditions for the studied mechanism.} \newnew{The parameter choices are plausible for a sub-set of disks, but should not be taken as representative for the entire population}. This will allow us to scrutinize the impact of viscous dissipation during binary-driven disk precession on inclination and obliquity distributions. \newnew{In Sect.~\ref{sect:discussion}, we will comment on the implications of this choice on the mechanism's ability to explain observed alignment trends in exoplanet-hosting binaries}.

\section{Observer-oriented coordinate system}
\label{sect:obs_oriented_system}

\begin{figure*}[t]
    \includegraphics[width = \linewidth]{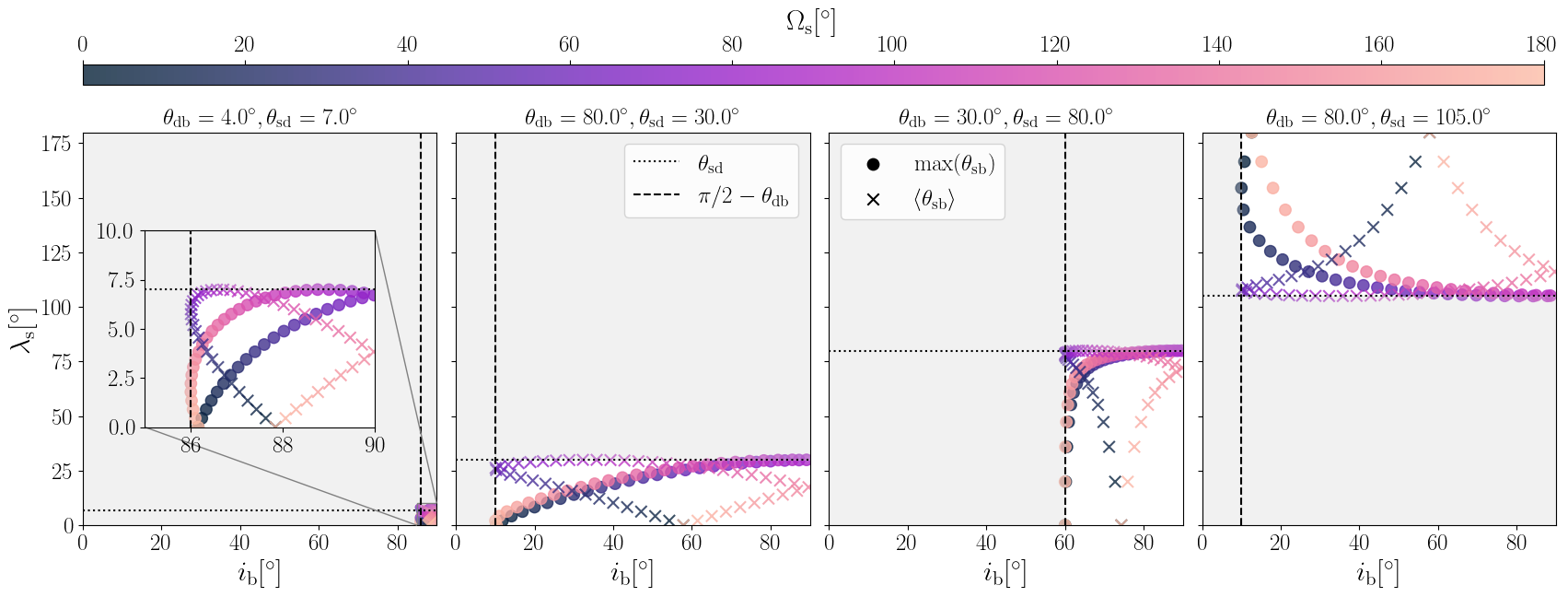}
    \caption{\new{Possible trajectories in $i_\mathrm{b}$-$\lambda_\mathrm{s}$ space for sets of $\theta_\mathrm{db}$, $\theta_\mathrm{sd}$ as the observer's longitude relative to the stellar spin $\Omega_\mathrm{s}$ changes (see Fig.~\ref{fig:angles}). From left to right, the four panels depict the following configurations: close-to-full alignment, close-to-polar binary and moderately misaligned spin, moderately misaligned binary and close-to-polar spin, and, close-to-polar binary and retrograde close-to-polar spin. Each trajectory is mapped out by 100 points spaced evenly from $\Omega_\mathrm{s} = 0$ to $\Omega_\mathrm{s} = \pi$. Their density thus can be interpreted as the probability $p(i_\mathrm{b}, \lambda_\mathrm{s})$. The true angles $\theta_\mathrm{db}$, $\theta_\mathrm{sd}$ define a forbidden region (gray shaded) which can never been observed. For given $\theta_\mathrm{db}$, $\theta_\mathrm{sd}$, the third invariant angle $\theta_\mathrm{sb}$ is geometrically constrained via $\mathrm{min}(\theta_\mathrm{sb}) = |\theta_\mathrm{db} - \theta_\mathrm{sd}| \leq \theta_\mathrm{sb} \leq \pi - |\pi - (\theta_\mathrm{db} + \theta_\mathrm{sd})| = \mathrm{max}(\theta_\mathrm{sb})$. Each panel shows two trajectories corresponding to $\mathrm{max}(\theta_\mathrm{sb})$ (circles) and $\langle \theta_\mathrm{sb}\rangle = [\mathrm{max }(\theta_\mathrm{sb}) - \mathrm{min }(\theta_\mathrm{sb})]/2$ (crosses). Note that the latter is an average of geometrically possible $\theta_\mathrm{sb}$ defined for visualization only and is unrelated to physical effects.}}
    \label{fig:projected_angle_visual}
\end{figure*}

To directly compare the results of this work with observations, it is useful to introduce an `observer-oriented' coordinate system \citep[compare to][]{Fabrycky2009}. In this \new{frame of reference}, the coordinate origin is at the location of the primary star, and the $\hat{\bm{z}}$ vector points towards the observer, \new{so that $\hat{\bm{x}} -\hat{\bm{y}}$ define the sky plane.} We define the line-of-sight inclinations of the three unit vectors as
\begin{align}
    i_\mathrm{d} & = \arccos(\hat{\bm{l}}_\mathrm{d} \cdot \hat{\bm{z}}), \\ 
    i_\mathrm{s} & = \arccos(\hat{\bm{s}}\cdot \hat{\bm{z}}), \\ 
    i_\mathrm{b} & = \arccos(\hat{\bm{l}}_\mathrm{b}\cdot \hat{\bm{z}}).
\end{align}
Figure~\ref{fig:angles} shows the set of unit vectors in this coordinate system. \new{Because the transiting planet forms in the disk, whose inclination we assume to evolve negligibly after disk dispersal, we fix $i_{\rm d}=90^\circ$, and are free to set the orientation of the sky plane so that the $y$-axis is parallel to $\hat {\bm {l}}_{\rm d}$. This defines the observer-oriented reference plane we will use from here on. Thus, $\theta_\mathrm{sd}$ is the planet's true obliquity (conventionally denoted as $\psi$), and $\theta_\mathrm{db}$ characterizes the  `orbit-orbit' angle, i.e. the true inclination between planetary orbit and binary orbit}.

\citet{Fabrycky2009} provide the angle transformations, i.e.
\begin{align}
\label{eq:lambda_s_calc}
    \tan \lambda_\mathrm{s} &= \frac{\sin \theta_\mathrm{sd} \sin \Omega_\mathrm{s}}{\cos \theta_\mathrm{sd} \sin i_\mathrm{d} + \sin \theta_\mathrm{sd}\cos \Omega_\mathrm{s} \cos i_\mathrm{d}}, \\
    \tan \lambda_\mathrm{b} &= \frac{\sin \theta_\mathrm{db} \sin \Omega_\mathrm{b}}{\cos \theta_\mathrm{db} \sin i_\mathrm{d} + \sin \theta_\mathrm{db}\cos \Omega_\mathrm{b} \cos i_\mathrm{d}},
\end{align}
\new{with $0 \leq \lambda_\mathrm{s}, \lambda_\mathrm{b} \leq \pi$.} Here, $\Omega_\mathrm{s}$ and $\Omega_\mathrm{b}$ are the longitudinal angles of stellar spin vector and binary orbit normal (see Fig.~\ref{fig:angles}). \new{For $i_\mathrm{d} = 90^\circ$, $\tan \lambda_\mathrm{s} = \tan \theta_\mathrm{sd}\Omega_\mathrm{s}$ and $\tan \lambda_\mathrm{b} = \tan \theta_\mathrm{db}\Omega_\mathrm{b}$}

\new{We also introduce an additional longitudinal angle, defined as the angle between $\hat {\bm s}$ and $\hat {\bm l}_{\rm b}$ after projection onto the $\hat {\bm x}$-$\hat {\bm z}$ plane:} 
\begin{align}
    \Omega_\mathrm{b} = \Omega_\mathrm{s} + \Delta \Omega,
\end{align}
\new{It} can be shown that (see Appendix~\ref{sec:rel_long}),
\begin{align}
\label{eq:rel_longitude}
\cos \Delta \Omega = \frac{\cos\theta_\mathrm{sb}}{\sin\theta_\mathrm{db}\sin\theta_\mathrm{sd}} - \frac{1}{\tan\theta_\mathrm{db}\tan\theta_\mathrm{sd}}.
\end{align}

We will take the azimuthal orientation of our system to be random \new{uniform}, which corresponds to $p(\Omega_\mathrm{s}) = \pi^{-1}$. \new{As the observables of interest are $\lambda_\mathrm{s}$ and $i_\mathrm{b}$ \citep[see][]{Rice2024}}, we are interested \new{in computing} the probability distributions $p(\lambda_\mathrm{s} | \theta_\mathrm{sd}, \Omega_\mathrm{s})$, and $p(i_\mathrm{b}| \theta_\mathrm{sb}, \theta_\mathrm{sd}, \theta_\mathrm{db}, \Omega_\mathrm{s})$ or $p(\cos i_\mathrm{b} | \theta_\mathrm{sb}, \theta_\mathrm{sd}, \theta_\mathrm{db}, \Omega_\mathrm{s})$. \new{The binary orbit inclination $i_\mathrm{b}$} relates to $\lambda_\mathrm{b}$ via \citep{Fabrycky2009}
\begin{align}
\label{eq:sinib_calc}
    \sin i_\mathrm{b} = \frac{\sin \theta_\mathrm{db}\sin\Omega_\mathrm{b}}{\sin \lambda_\mathrm{b}},
\end{align}
while the inclination of the spin vector relative to the sky plane is
\begin{align}
    \sin i_\mathrm{s} = \frac{\sin \theta_\mathrm{sd}\sin\Omega_\mathrm{s}}{\sin \lambda_\mathrm{s}}.
\end{align}
\new{Figure~\ref{fig:projected_angle_visual} visualizes possible trajectories in $i_\mathrm{b}$-$\lambda_\mathrm{s}$ space (observed binary inclination and stellar obliquity), for four sets of $\theta_\mathrm{db}$ and $\theta_\mathrm{sd}$ (`true' binary inclination and stellar obliquity). For all $\Omega_\mathrm{s}$, $i_\mathrm{b} \geq \pi/2 - \theta_\mathrm{db}$. For stellar spins in prograde rotation relative to $\hat{\bm{l}}_\mathrm{d}$, i.e. if $\theta_\mathrm{sd} < \pi/2$, then $\lambda_\mathrm{s} \leq \theta_\mathrm{sd}$ always. For retrograde stellar spins with $\theta_\mathrm{sd} > \pi/2$, $\lambda_\mathrm{s} \geq \theta_\mathrm{sd}$.}

\subsection{Initial configuration}
\label{sect:initial_config}

The three main mechanisms for forming stellar binary systems are disk fragmentation, turbulent fragmentation, and dynamical capture, which are associated with different expectations for the primordial inclination between binary orbit and stellar spin orbit.While dynamical capture results in largely random inclinations, disk fragmentation is expected to initialize the true binary inclination as preferentially aligned. Since those two mechanisms are typically invoked for the formation of very wide ($> 10^4$ AU) \citep{Tokovinin2017} and relatively close ($< 500$ AU) \citep{Krumholz2007, Tobin2018} binaries, respectively, we instead focus on turbulent fragmentation, which is shown to manufacture binaries with orbital separations throughout our range of interest, i.e. \new{$400 < a_\mathrm{b} < 1000$} AU \citep{Padoan2002,Bate2018,Lee2017}, and with, to zeroth order, random orientations.

We thus assume the orientation of stellar spin angular momentum vector to be random uniform. The corresponding probability density function for the angle is given by $p[\theta_\mathrm{db}(t=0)] \propto \sin{|\theta_\mathrm{db}|}$. Note that, in our model, retrograde \new{binary} orbits behave identically to prograde orbits. We will focus our calculations on initial misalignments $0^\circ \leq \theta_\mathrm{db} \leq 90^\circ$.

\subsection{Example model}
\label{sect:example_evolution}

\begin{figure*}[t]
    \centering
    \includegraphics[width = \linewidth]{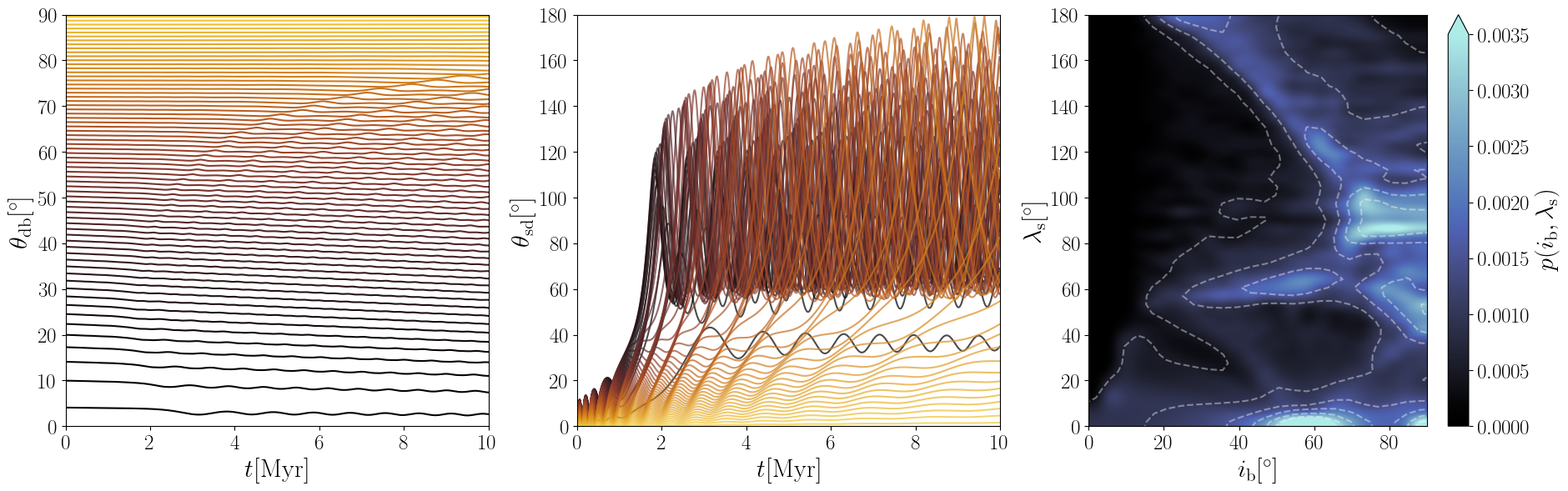}
    \caption{Evolution of disk-binary angle $\theta_\mathrm{db}$ (left panel) and stellar spin- disk angle $\theta_\mathrm{sd}$ (center panel) for \new{models} initialized at different inclinations.  The probability map $p(i_\mathrm{b}, \lambda_\mathrm{s})$ is shown in the right panel \new{(see} Sect.~\ref{sect:obs_oriented_system}). The depicted models use our fiducial parameter set for an equal-mass binary $M_* = M_\mathrm{b} = M_\odot$ with a binary separation of $a_\mathrm{b} = 600 $ AU (see Tab.\ref{tab:parameters}).}
    \label{fig:fiducial_evolution}
\end{figure*}

We calculate the evolution of the vectors $\hat{\bm{l}}_\mathrm{d}$ and $\hat{\bm{s}}$ over $t_\mathrm{l} = 10^7$ yr, a fiducial value for the lifetime of the gas disk consistent with expectations from photoevaporation models \citep{Ercolano2009, Owen2012}. \new{We then take the final values of $\theta_{\rm sd}$, $\theta_{\rm sb}$, and $\theta_{\rm db}$ as} inputs for Eqs.~\eqref{eq:lambda_s_calc} and \eqref{eq:sinib_calc} to calculate probability density functions for the angles $\lambda_\mathrm{s}$ and $i_\mathrm{b}$. Note, that resonance crossing can put the system into a state of oscillating $\theta_\mathrm{sd}$ with amplitudes up to tens of degrees \citep[e.g.,][]{Lai2014}. \new{The phase of this oscillation at $t = t_\mathrm{l}$ and the associated final value for the spin orbit angle are sensitive to the initial condition, requiring sufficiently fine sampling of initial binary inclinations $\theta_\mathrm{db}$.}

Fig.~\ref{fig:fiducial_evolution} shows the evolution of the invariant angles $\theta_\mathrm{db}$ and $\theta_\mathrm{sd}$ (left and center panel respectively), where color corresponds to the initial $\theta_\mathrm{db}$, for our set of default parameters depicted in Table~\ref{tab:parameters}, as well as $M_* = M_\mathrm{b} = M_\odot$, at a separation of $a_\mathrm{b} = 600$ AU. As studied in \citet{Zanazzi2018torquedamp, Zanazzi2018}, the depletion of disk material causes $\tilde{\omega}_\mathrm{sd}$ to fall below $\tilde{\omega}_\mathrm{db}$, triggering a resonance crossing. The right panel of Fig.~\ref{fig:fiducial_evolution} shows the probability density map $p(i_\mathrm{b}, \lambda_\mathrm{s})$ in arbitrary units. Most initial alignments lead obliquity excitation \new{into a state of oscillating $\theta_\mathrm{sd}$}. \new{Many systems end up with a retrograde spin-orbit angle $\theta_\mathrm{sd} > \pi/2$, which results in observed trajectories in $i_\mathrm{b}$-$\lambda_\mathrm{s}$- space with $\pi > \lambda_\mathrm{s} > \pi/2$ (see Fig.~\ref{fig:projected_angle_visual}). Systems not experiencing significant excitation of obliquity are mostly those initialized with close-to polar $\theta_\mathrm{db}$. These correspond to $i_\mathrm{b}$-$\lambda_\mathrm{s}$- space trajectories similar to the one shown in the second from the left panel in Fig.~\ref{fig:projected_angle_visual}, thus populating the bottom part of the probability density map $p(i_\mathrm{b}, \lambda_\mathrm{s})$ in the right panel of Fig.~\ref{fig:fiducial_evolution}.} While we recover the result in \citet{Zanazzi2018torquedamp} -- namely, that viscous damping for the chosen set of parameters is too weak to fully drive individual disks towards alignment with the binary plane -- the population of disks as a whole shifts away from an isotropic distribution towards a more aligned distribution. We will elaborate on this finding further in the following section.

\section{Results}
\label{sect:results}

\begin{figure*}[t]
    \centering
    \includegraphics[width = \linewidth]{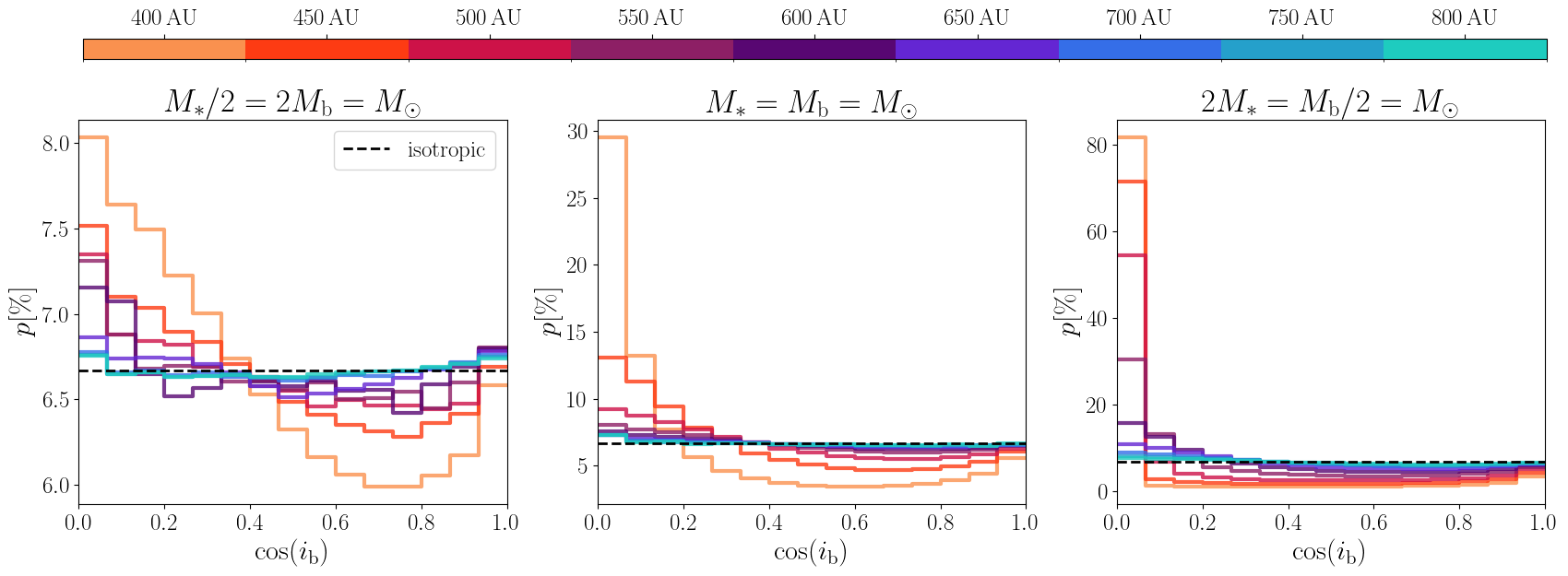}
    \caption{Probability of measuring $\cos(i_\mathrm{b})$ for binary separations between 400 AU and 800 AU. Different panels correspond to different binary configurations: circum-primary disk with less massive stellar companion (left), equal mass binary (center), and circum-secondary disk with more massive stellar companion (right).}
    \label{fig:cosi_for_binarysep}
\end{figure*}

\begin{figure*}[t]
    \centering
    \includegraphics[width = \linewidth]{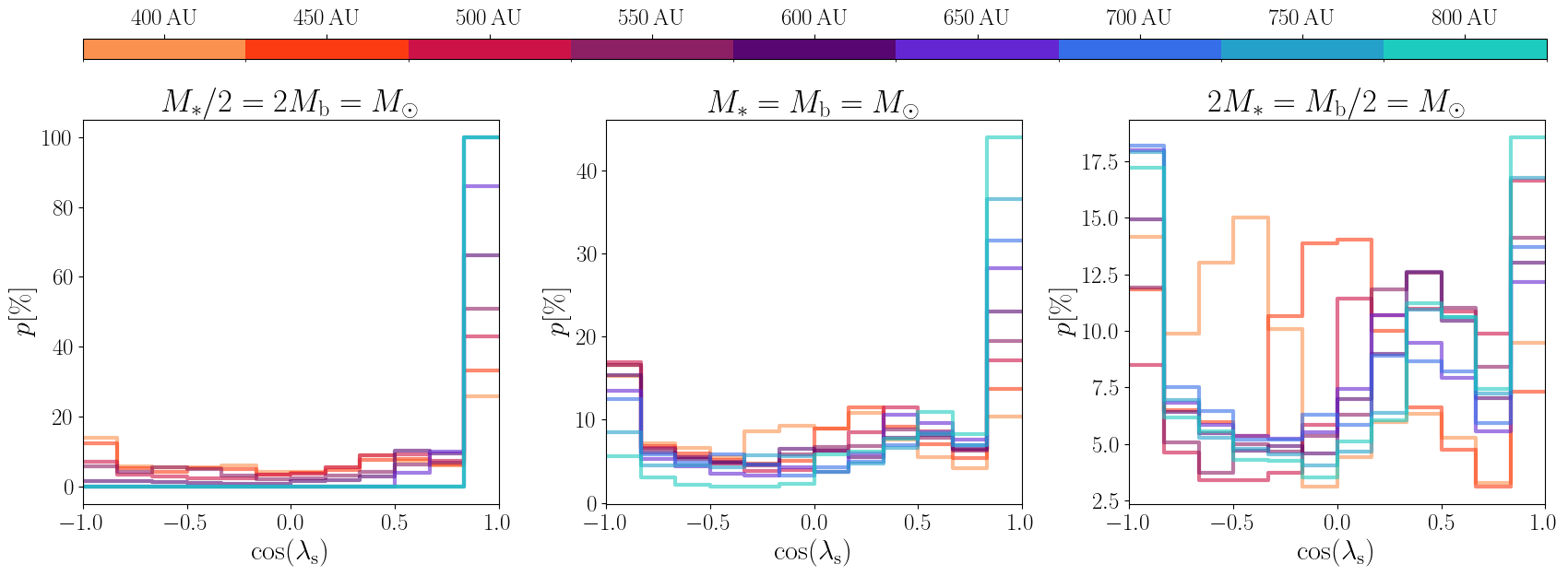}
    \caption{Same as Fig.~\ref{fig:cosi_for_binarysep} but for sky-projected obliquity $\cos(\lambda_\mathrm{s})$. The predicted distribution\new{, specifically for equal-mass-ratio binaries, is} broadly consistent with observed trends in \citet{Dong2023, Siegel2023} \new{and \citet{Rice2024}}.}
    \label{fig:lam_for_binarysep}
\end{figure*}

\begin{figure}
    \centering
    \includegraphics[width=\linewidth]{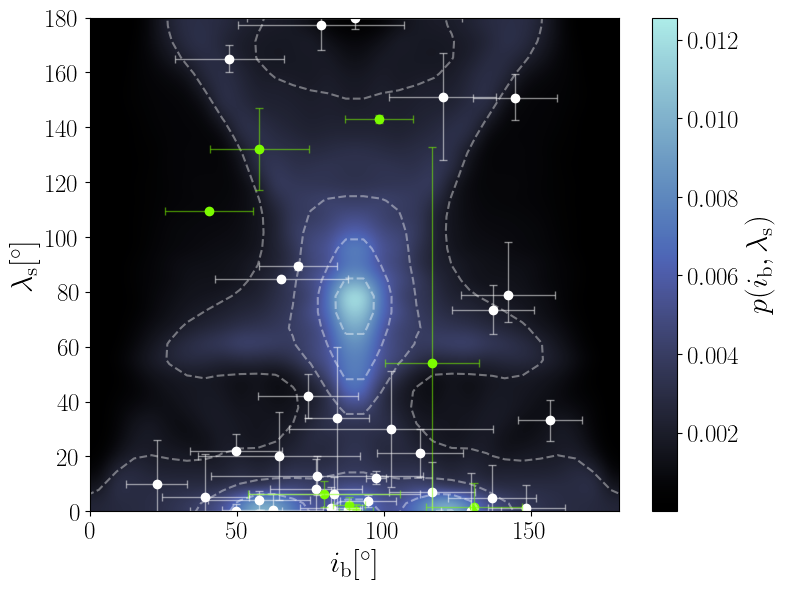}
    \caption{\new{Cumulative probability map of equal-mass-ratio models shown in the center panels of Figs.~\ref{fig:cosi_for_binarysep} and \ref{fig:lam_for_binarysep}. In contrast to the right panel of Fig.~\ref{fig:fiducial_evolution}, this probability map covers $0^\circ \leq i_\mathrm{b} \leq 180^\circ$, achieved by mirroring at $i_\mathrm{b} = 90^\circ$. We also set the model prediction side-by-side with the data of the `strict' sample of planet-hosting binaries with obliquity measurements in \citet{Rice2024}. White dots belong to systems with binary separations that fall within $400 \,\mathrm{AU} < a_\mathrm{b} < 800\,\mathrm{AU}$, whereas green dots are systems either inside or outside this range.}  }
    \label{fig:Rice2024_comparison}
\end{figure}

We aim to characterize the maximum binary separation at which \new{dissipative precession} significantly causes deviations from isotropy in the final distribution of line-of-sight binary inclination $i_\mathrm{b})$ and sky-projected stellar obliquity $\lambda_\mathrm{s}$, as well as the expected variations for circum-primary versus circum-secondary planets in binary systems with different mass ratios. For this purpose, Figs.~\ref{fig:cosi_for_binarysep} and \ref{fig:lam_for_binarysep} show the probabilities of measuring $\cos(i_\mathrm{b})$ and $\lambda_\mathrm{s}$, respectively, for three binary mass configurations across binary separations ranging from 400-800 AU.

\subsection{Equal-mass binary}
\label{sect:results_equalmass}

The damping rate dramatically weakens with binary separation, such that for typical disk parameters, even moderately wide binaries have damping timescales that exceed expected disk lifetimes (see Fig.~\ref{fig:cosi_for_binarysep}). Nonetheless, a population of disks subject to this `weak' damping can still have a distribution of sky-projected angles that deviate from random orientations. To quantify the maximum binary separation for which we reasonably expect this to occur, we first consider equal-mass binaries with $M_* = M_\mathrm{b} = M_\odot$.

The center panel of Fig.~\ref{fig:cosi_for_binarysep} demonstrates that, for the fiducial setup, up until a binary separation of $\sim$ 500 AU our model predicts an orientation distribution that deviates from isotropy -- specifically, an excess of aligned systems and a lack of misaligned systems. Despite parameter choices that are relatively favorable for alignment, this cut-off is short of the boundary between preferentially aligned and randomly oriented systems, which \citet{Christian2022align} reported to be located between 600 and 800 AU. Moreover, in our study, the inclinations of systems initialized in a near-polar configuration ($\cos i_{\rm b} \sim1.0$) remain largely unaffected by the inclination damping studied in this paper. While this feature has not been identified in observations \citep{Christian2022align, dupuy2022orbital}, where the dearth of misaligned systems may persist at polar orientations, it may be a useful indicator for efficient dissipation if seen in future observations that incorporate larger sample sizes. 

\new{For these reasons}, the herein studied mechanism alone can likely not explain the observed orientations of planet-hosting binaries, but instead needs to be accompanied by an additional process such as primordial alignment of the disk or post planet formation reorganization. We expand on this in Sect.~\ref{sect:additional_alignment_pathways}.

Obliquity excitation is less sensitive to semi-major axis (center panel of Fig.~\ref{fig:lam_for_binarysep}) and still occurs up to $a_\mathrm{b} = 800$ AU, although for a smaller fraction of systems than at smaller binary separations.   At larger binary separations, secular resonance ($\tilde \omega_{\rm sd} \sim \tilde \omega_{\rm db}$) occurs at later times, prolonging the excitation of $\lambda_{\rm s}$ to timescales longer than those given by Eq.~\eqref{eq:t_upper}.  For sufficiently large $a_{\rm b}$, the binary's torque becomes too weak to excite $\lambda_{\rm s}$.  This is why the sky projected obliquity $\lambda_{\rm s}$ tends to decrease -- or $\cos \lambda_{\rm s}$ tends to increase -- as $a_{\rm b}$ gets larger.

Overall, the predicted distribution qualitatively agrees with recently reported obliquity distributions, both specifically for planet-hosting binaries \citep{Rice2024} and also across the full census of systems with spin-orbit measurements \citep{Albrecht2021, Dong2023, Siegel2023}. For large binary separations, most systems do not experience obliquity excitation, and the stellar spin and disk orbit normal remain aligned. Misaligned systems are expected for a wider range of initial orientations at smaller binary separations. \citet{Albrecht2021} report an overabundance of systems that are perpendicular ($\cos \lambda_s \sim 0$) in their spin-orbit orientations (although see also \citet{Siegel2023} and \citet{Dong2023}, which show that this peak is not significant in a hierarchical Bayesian framework). A broad peak near $\cos \lambda_s = 0$ may be consistent with our \new{model} if \new{resonant crossing preferentially excites obliquities to near-polar configurations} \citep[see also][]{Vick2023}. \new{Indeed, the systems with significant spin-orbit misalignment tend to be in a state of high-amplitude oscillation of $\theta_\mathrm{sd}$ around a near-polar value (see Fig.~\ref{fig:fiducial_evolution}). The herein utilized parameter set, however, does not damp these oscillations to the mid-point value sufficiently fast, ultimately resulting in a more spread-out distribution of $\cos(\lambda_\mathrm{s})$ in Fig.~\ref{fig:lam_for_binarysep}.}

\new{The cumulative probabilities $p(i_\mathrm{b}, \lambda_\mathrm{s})$ of all models with equal stellar mass ratio are shown in Fig.~\ref{fig:Rice2024_comparison} in comparison to the observed joint obliquity -- binary inclination -- distribution reported in \citet{Rice2024}. For this purpose, Fig.~\ref{fig:Rice2024_comparison} mirrors the probability map at $i_\mathrm{b} = 90^\circ$. We only show the 40 systems in the `strict' sample from \citet{Rice2024}, out of which 8 have binary separation within 400 and 800 AU (red dots). The fiducial model and the (sparse) data have the following trends in common: most systems appear close to spin-orbit alignment, including a pile-up of fully aligned systems with a wide range of allowed obliquities. We are not concerned that the shape of the probability map does not accurately track the data points from \citet{Rice2024}, as (1) the data of planet-hosting binaries with obliquity measurements are sparse, (2) obliquity $\lambda_\mathrm{s}$ (arguably unlike binary inclination $i_\mathrm{b}$) can be affected by a wide variety of other dynamical effects that operate during or after the protoplanetary disk phase \citep[e.g.,][]{Anderson2018, Millholland2019, Li2021, Vick2023}, and (3) the model was (intentionally) not set up as a population synthesis, but to highlight potential observable implications of the physics of dissipative precession in the circumstellar disk. Such a population synthesis would need to carefully consider the types of systems that underlie the observations, explore a larger parameter space, and involve some treatment of post-disk dynamical mechanisms that affect observables. We defer this to future work.}

\subsection{Hierarchical binary}
\label{sect:unequal_massratio}

We showed that disks in equal-mass-ratio binaries, on a population level, can noticeably experience alignment. Yet, tension remains: the viscous dissipation mechanism, even with optimistic model parameters, becomes inefficient for exceedingly wide binaries ($a_\mathrm{b} \gtrsim 600$ AU). Consequently, \new{the study so far is an insufficient argument for} alignment of exoplanet-hosting binaries \new{via} viscous dissipation during the binary-driven recession of protoplanetary disks, \new{and} other potential observable consequences \new{must be examined}. The mass ratio dependence of Eq.\eqref{eq:damping_rate_final} provides this opportunity.

We consider a hierarchical binary: specifically, a system with mass ratio of $M_1/M_2 = 4$. Figures~\ref{fig:cosi_for_binarysep} and \ref{fig:lam_for_binarysep} show corresponding orientation distributions in inclination and obliquity for the circum-primary systems (disk host star mass $M_* = 2 M_\odot$ and companion star mass $M_\mathrm{b} = 0.5 M_\odot$) in the left panels, and for circum-secondary systems  (disk host star mass $M_* = 0.5 M_\odot$ and companion star mass $M_\mathrm{b} = 2 M_\odot$) in the right panels.

Our key finding from this exercise is that, for circum-primary systems, the final inclination distribution \new{more closely resembles an isotropic distribution than for equal-mass-ratio binaries, even at small projected separations of}  $a_\mathrm{b} \sim 400$ AU. For circum-secondary systems, by contrast, the inclination distribution is skewed up to $a_\mathrm{b} \sim 650$ AU, comparable to the expected distribution for equal-mass ratio systems (see center panel of Fig.~\ref{fig:cosi_for_binarysep} and Sect.~\ref{sect:results_equalmass}).

Because a more massive companion increases the magnitude of the binary's torque, spin-orbit misalignments occur more often in circum-secondary than circum-primary systems. Few circum-primary systems experience obliquity excitations, and in general only within binary separations of $\lesssim 500$ AU. On the other hand, circum-secondary systems \new{almost always} have misaligned stellar obliquities, even at larger binary separations.

\section{The Mass Dependence of Alignment in Observed Systems}
\label{sect:obs_data}

\begin{figure*}
    \centering
    \includegraphics[width = \linewidth]{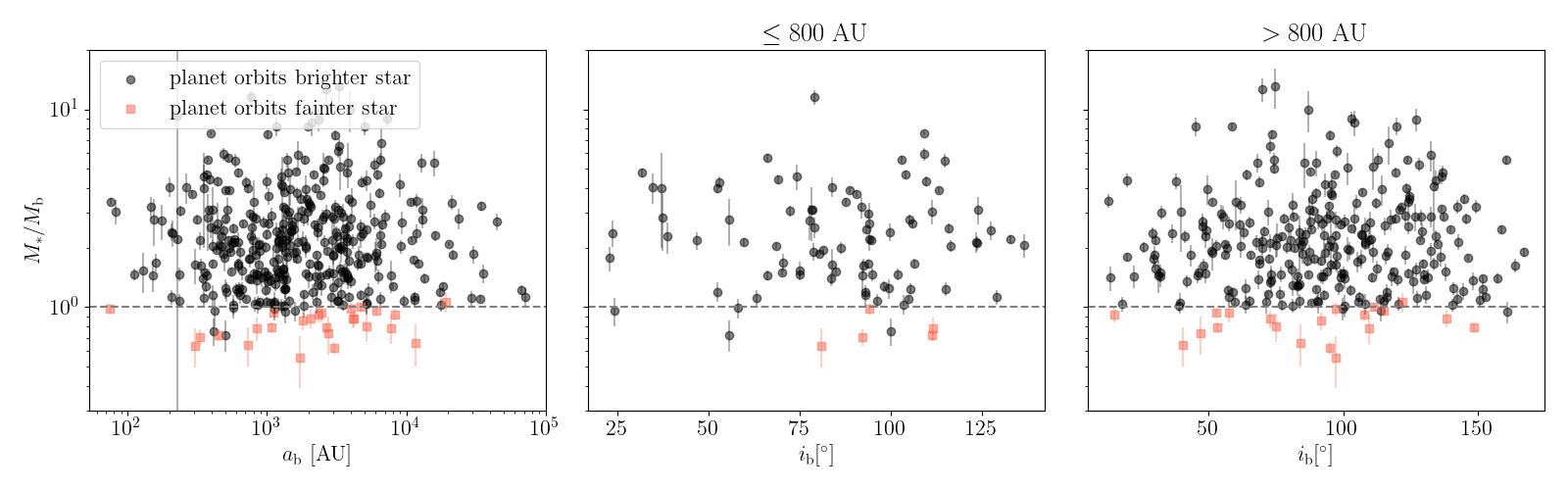}
    \caption{Stellar mass ratio vs binary separation (left panel) and mass ratio vs sky-projected inclination (center and right panel) for the sample of planet-hosting binaries selected and curated by \citet{Christian2023inprep}. We delineate the sample into two subsets lying inside and outside a binary separation of $800$ AU, as this is approximately where \citet{Christian2022align} find evidence for preferential alignment.}
    \label{fig:data_scatter}
\end{figure*}

\begin{figure*}
    \centering
    \includegraphics[width = \linewidth]{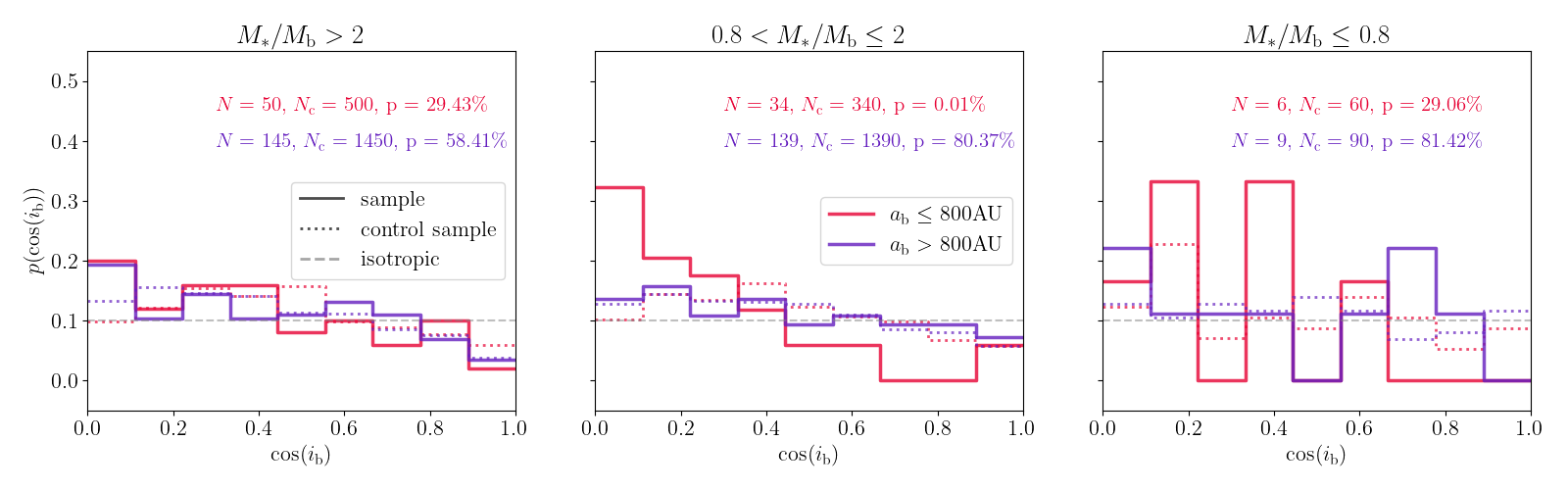}
    \caption{Observed distributions of sky-projected inclinations for circum-primary planets (left panel), near-equal mass ratio systems (center panel), and circum-secondary planets (right panel). The data are split into two semi-major axis bins inside or outside 800 AU. The control sample is shown in dotted lines. The fact that the control sample deviates from isotropy is indicative of systematic biases as discussed in \citet{Christian2022align}. We also show the $p$-value resulting from a KS-test for each population. }
    \label{fig:data_hist}
\end{figure*}

The \new{damping via dissipative precession} depends on stellar mass ratio, as discussed in Sects.~\ref{sect:results_equalmass} and \ref{sect:unequal_massratio}. It is thus of immediate interest to investigate a potential correlation between stellar mass ratio and orbital orientations in observed exoplanetary populations. 

\subsection{Sample selection and data curation}

Our sample is drawn from the intersection of systems observed with one or more transiting exoplanet candidates in the \textit{Gaia} DR2-based TESS Input Catalog (TIC); two resolved stars in \textit{Gaia} DR3; and a sufficiently robust astrometric solution/low probability of chance alignment as quantified by \citet{ElBadry2021}. After removal of false positives, the sample includes 398 candidate exoplanet-hosting binaries. We also leverage the control sample of binary systems without observed transiting exoplanets from \citet{Christian2023inprep} in our analysis. This allows us to filter for selection effects. Stellar masses for both populations were obtained using a combination of empirical mass-luminosity relations and stellar isochrone fitting. We refer to \citet{Christian2023inprep} and \citet{Christian2022align} for further details on the data selection and curation, as well as the control sample.

For all of the binaries in the sample, the separation is less than 21 arcseconds on the sky, which is the size of a single TESS pixel. It is therefore difficult in general to determine which of the two stars the transiting exoplanet is orbiting. While in some cases the TESS data can be used to identify the true host star by fitting the telescope's pixel response function to the in/out-of-transit difference image, in many other cases (especially for the closest binaries that are most likely to be aligned) we cannot confidently identify which star is the host unless there is ground-based follow-up data. 

In order to proceed, we assume that all planets in our sample orbit the host stars indicated in the TESS Objects of Interest catalog \citep{Guerrero2021ApJS}. Effectively, this means that we assume planets all orbit the more massive (brighter) star unless there is compelling evidence to the contrary (either from TESS centroid analysis or ground-based follow-up observations). This is generally a good assumption because of the biases of the transit method, which is more sensitive to deeper transit signals that are more easily observable from the brighter star of a blended signal \citep{Ciardi2015ApJ}.

Figure~\ref{fig:data_scatter} shows stellar mass ratio vs binary separation for the sample (left panel), as well as mass ratio vs measured inclination (center and right panel). There are 15 systems with a mass ratio of $M_*/M_\mathrm{b} \leq 0.8$, only four of which have semimajor axes less than 800 AU. This is too few for any meaningful statistical statements. On the other hand, there is a large number of systems that we can broadly assign to the equal mass ratio and circum-primary categories, defined here as $0.8 < M_*/M_\mathrm{b} \leq 2$ and $M_*/M_\mathrm{b} > 2$, respectively. The data thus present an ideal proving ground to test the significance of the difference in the inclination distribution between equal-mass ratio and circum-primary systems (see Fig.~\ref{fig:cosi_for_binarysep}).

\subsection{Orientation distribution as a function of stellar mass ratio}

Figure~\ref{fig:data_hist} shows the $\cos(i_\mathrm{b})$ distribution of the sample, separated by stellar mass ratio and binary separation. We overplot equivalent distributions for the control sample, which includes no transiting planets, yet likewise deviates from isotropy due to systematic biases \citep[see][for a discussion]{Christian2022align}. It is thus more appropriate to compare the planet-hosting sample to the control sample rather than the line of isotropy which is drawn in Fig.~\ref{fig:data_hist} for reference. 

Equal mass ratio systems (center panel of Fig.~\ref{fig:data_hist}) exhibit a trend consistent with that published in \citet{Christian2022align} and \citet{Christian2023inprep}: exoplanet-hosting binaries with semi-major axes greater than $\sim$ 800 AU are consistent with the (almost isotropic) control sample, whereas those with semi-major axes less than $\sim$ 800 AU visibly display preferential alignment. This is confirmed by a two-sample Kolmogorov-Smirnov (KS) test, where the resulting $p$-value characterizes the probability that the two samples originate from the same underlying distribution. The null hypothesis typically is rejected if $p < 5\%$. Indeed, the two sample KS-test yields $p$-values of $0.01 \%$ and $80.4 \%$ for the equal mass ratio populations inside and outside 800 AU respectively, when comparing to the associated control sample population. 

The high-mass ratio systems are distributed -- at least by eye -- more isotropically. Indeed, a KS test with the corresponding control sample yields $p$-values of $29.4 \%$ and $58.4 \%$ for the high mass ratio populations inside and outside 800 AU, respectively. The KS test therefore cannot reject the null hypothesis that exoplanet-hosting high-mass ratio binaries are oriented in a similar fashion to the binaries in the control sample.  

To conclude, while the observational data are relatively sparse, they are broadly consistent with the theoretical predictions made in Sect.~\ref{sect:unequal_massratio} and Fig.~\ref{fig:cosi_for_binarysep}.

\section{Additional \new{alignment pathways}}
\label{sect:additional_alignment_pathways}

In our analysis of exoplanet-hosting binary systems, we observed characteristics consistent with alignment through viscous dissipation. However, our findings also indicate that viscous dissipation alone may not fully account for the current orientation distribution of these systems (see Sect.~\ref{sect:results_equalmass}). While present-day data is ambiguous, if polar systems are indeed rare, this phenomenon could not be adequately reproduced by dissipative precession -- at least as treated in the models presented here. 

In this section, we therefore discuss two other possibilities -- namely, preferential primordial alignment and removal of highly inclined systems through the Kozai-Lidov mechanism -- alongside their prospects to, jointly with \new{dissipative precession}, sculpt the orbital configurations of planet-hosting binaries.

\subsection{Primordial alignment \& magnetic torques}

Our models assumed an initial configuration in which the spin and disk angular momentum vectors were aligned but randomly oriented relative to the binary angular momentum vector, leading to a probability density function for the invariant disk -- binary inclination $\theta_\mathrm{db}$ given by $\sin |\theta_\mathrm{db}|$. As discussed in Sect.~\ref{sect:initial_config}, this choice aims to reflect turbulent fragmentation, where the binary components are assumed to form independently and with random orientations, as the mode of star formation examined within this work. 

Numerical investigations, however, suggest that, even for wide binary formation via turbulent fragmentation, a dearth of polar systems is expected \citep[e.g.,][]{bate2010chaotic, Bate2018}. Physically, this is a consequence of the conservation of the angular momentum of the contracting cloud core, which, even in the presence of strong turbulence, will lead to a non-isotropic orientation of the angular momentum vectors of individual stars. 

In addition, accretion of ambient gas onto the binary components acts to align the binary inclination \citep{Bate2018}. Torques induced by gas accretion can also hinder excitation of spin-orbit misalignment \citep{Lai2014}, thus affecting expected obliquity distributions beyond what is discussed in our paper. The obliquity distribution can, furthermore, be impacted by magnetic torques. \citet{Lai2011, Lai2014} show that magnetic torques contribute to precession rates and thus to the excitation of $\theta_\mathrm{sd}$. On the other hand, magnetic torques also can act to (re-)align stellar spin with the disk, specifically for low mass stars \citep{Spalding2015}. As magnetic effects generally do not affect the mutual disk-binary inclination -- the focus of this paper -- we do not include them in this study.

\subsection{Kozai-Lidov oscillations in polar configurations}
\label{sect:kozai_planet}

Kozai-Lidov oscillations are suppressed as long as the disk self-gravity is strong enough to force eccentric orbits in apsidal precession (see Sect~\ref{sect:disk_kozai}). On the other hand, once the protoplanetary disk  evaporates, and provided the binary inclination is sufficiently large, the Kozai-Lidov mechanism can freely operate, in the process affecting the orbital configurations of the planets within the system. Specifically, this can lead to a distribution where polar configurations are removed \citep{Naoz2012}. Planet-planet interactions become more frequent in systems undergoing Kozai-Lidov cycles \citep{Malmberg2007}, which, for highly inclined systems, may lead to scattering and thus the removal of such systems from the observed orientation distributions.

The Kozai-Lidov mechanism may, therefore, reconcile the inclination distributions produced by this paper's model (e.g., Fig.~\ref{fig:cosi_for_binarysep}) with the observed distributions in Fig.~\ref{fig:data_hist} or in \citet{Christian2022align, Rice2024}, where polar systems are uncommon.

\section{Conclusions}
\label{sect:discussion}

In this paper, we investigated the prospects for viscous dissipation -- occurring during precession of the protoplanetary disk driven by an inclined binary companion -- to sculpt resulting orbital configurations. \new{Physically, the disk's mass loss is associated with a resonant crossing of precession frequencies which excites the system's spin-orbit angle (Sect.~\ref{sect:equations_of_motion}). At the same time, the orbit-orbit angle is damped as the disk dissipates energy in an effort to resist twisting by the torque from the binary companion (Sect.~\ref{sect:viscous_diss_and_disk_model}).} The \new{sky-projected} inclination and obliquity distributions predicted by \new{this model} are broadly consistent with observed trends \new{(Sects.~\ref{sect:obs_oriented_system} and \ref{sect:results})}. Specifically, \new{if} dissipative precession \new{can operate effectively, it may} offer a joint explanation for:
\begin{itemize}
    \item The preferential \new{population-level} alignment of binaries that host transiting exoplanets perpendicular to the sky plane, and the decrease in alignment for increasing binary separations \citep{dupuy2022orbital, Christian2022align}.
    \item The novel finding that this preferential alignment \new{appears to be} more pronounced for binaries with close-to equal stellar mass ratios compared to circum-primary planets within high mass ratio binaries, where it may not exist at all \new{(Sect.~\ref{sect:obs_data})}.
    \item The stellar obliquity distribution, which consists of mostly aligned systems, with a second population of significantly misaligned or isotropically distributed systems \citep{Albrecht2021, Dong2023, Siegel2023, Rice2024}.
\end{itemize}
\new{An important caveat is that the damping rate due to viscous dissipation during binary-driven disk precession weakens with increasing binary separation $a_\mathrm{b}$ and decreasing disk extent $r_\mathrm{out}$, viscosity $\alpha$ and binary mass ratio $M_\mathrm{b}/M_*$ via Eq.~\eqref{eq:damping_rate_final}. Observed protoplanetary disks show a large diversity in these properties (see Sect.~\ref{sect:diskmodel_context}), and consequently, dissipative precession is expected to only be efficient for the sub-set of systems adequately described with parameters around those chosen by our fiducial model ($r_\mathrm{out} \gtrsim 100$ AU, $\alpha \gtrsim 0.05, M_\mathrm{b}/M_* \gtrsim 1$).} Additional constraints on disk properties, specifically their radial extent and their viscosity, will help to pinpoint the role of alignment via dissipative precession.

Our model presupposes that star formation results in a spherically isotropic distribution of initial inclinations. Such a distribution would leave behind  a population of polar systems seemingly absent in the observational data. However, this discrepancy can be mitigated by viscous dissipation during binary-driven recession juxtaposed with even small degrees of primordial alignments, removal of polar systems via Kozai-Lidov oscillations, or both \new{(Sect.~\ref{sect:additional_alignment_pathways})}. \new{While the presented model can reproduce the broad, qualitative trends \newnew{under favorable conditions}, we suspect that a full sophisticated population synthesis, which} jointly models these effects \new{and covers the full parameter space of appropriate disk properties, is required to match the observed, joint - obliquity - binary inclination - distribution in \citet{Rice2024}. As the observational sample of obliquity measurements in exoplanet-hosting binaries grows, so will the appeal of such an undertaking.} 

Finally, our work demonstrates that the distributions of planet orbit orientations and stellar obliquities in binaries may be dependent on stellar mass ratio, thus highlighting that planet-hosting binaries are a unique window into the planet formation process. In this specific context, additional demographic constraints on planets in high mass-ratio binary systems, in particular, difficult-to-identify circum-secondary planets, may be \new{exceptionally} decisive in diagnosing the predominant mechanism that drives exoplanets towards alignment with their stellar binary.

\section{Acknowledgements}
\label{sec:acknowledgements}
\new{We thank the referee for insightful comments.} K.G. thanks Jeremy Smallwood, Mark Dodici, Yubo Su, Jiayin Dong, Callum Fairbairn and Tiger Lu for insightful discussions. M.R. acknowledges support from Heising-Simons Foundation Grants \#2023-4478 and \#2023-4655, NASA grant GR122985/AWD0011072, and Oracle for Research grant No. CPQ-3033929. JJZ was supported by a 51 Pegasi b Heising-Simons Fellowship.

\software{\new{NumPy \citep{Harris2020}, Matplotlib \citep{Hunter2007}, CMasher \citep{vanderVelden2020}}}.

\appendix

\section{Calculation of the relative longitude $\Delta \Omega$}
\label{sec:rel_long}

Consider the set of vectors and angles depicted in Fig.~\ref{fig:angles}. We denote the projections of $\hat{\bm{s}}$ and $\hat{\bm{l}}_\mathrm{b}$ onto the $x-z$ plane as $\hat{\bm{s}}_{xz}$ and $\hat{\bm{l}}_{\mathrm{b},xz}$ respectively, and the projections onto the $y$-axis as $\hat{\bm{s}}_{y}$ and $\hat{\bm{l}}_{\mathrm{b},y}$ respectively.

$\hat{\bm{s}}$ and $\hat{\bm{l}}_\mathrm{b}$ create a triangle, where the remaining side $\overline{SB}$ is opposite to the angle $\theta_\mathrm{sb}$. Since $\hat{\bm{s}}$ and $\hat{\bm{l}}_\mathrm{b}$ are unit vectors, the law of cosines implies
\begin{align}
    \overline{SB}^2 = 2(1-\cos\theta_\mathrm{sb}).
\end{align}
We continue by considering the plane $P$ parallel to the $y$-axis that contains $\overline{SB}$. At $y = 0$, the plane also contains the remaining side $\overline{Y}$ of the triangle constructed by $\hat{\bm{s}}_{xz}$ and $\hat{\bm{l}}_{\mathrm{b},xz}$.  $\overline{Y}$ is opposite to the relative longitude $\Delta \Omega$, so we apply the law of cosines again as
\begin{align}
    \cos \Delta \Omega = \frac{|\hat{\bm{s}}_{xz}|^2 + |\hat{\bm{l}}_{\mathrm{b},xz}|^2 - \overline{Y}^2}{2|\hat{\bm{s}}_{xz}||\hat{\bm{l}}_{\mathrm{b},xz}|},
\end{align}
where $|\hat{\bm{s}}_{xz}| = \sin\theta_\mathrm{sd}$ and $|\hat{\bm{l}}_{\mathrm{b},xz}| = \sin\theta_\mathrm{db}$ since $\hat{\bm{s}}$ and $\hat{\bm{l}}_\mathrm{b}$ are unit vectors, and $\hat{\bm{l}}_\mathrm{d}$ is aligned with the $y$-axis. We consider a right-angled triangle in $P$, that has hypotenuse $\overline{SB}$ and the legs $\overline{Y}$ and $\overline{D}$, such that $\overline{SB}^2 = \overline{Y}^2 + \overline{D}^2$. Here, $\overline{D}$ is parallel to $\hat{\bm{l}}_\mathrm{b}$ and the $y$-axis, and can be defined as $\overline{D} = |\hat{\bm{s}}_{y}| - |\hat{\bm{l}}_{\mathrm{b},y}|$. These projections have magnitudes of $|\hat{\bm{s}}_{y}| = \cos \theta_\mathrm{sd}$ and $|\hat{\bm{l}}_{\mathrm{b},y}| = \cos \theta_\mathrm{db}$. 

We can therefore write
\begin{align}
    \overline{Y}^2 = \overline{SB}^2 - \overline{D}^2 = \overline{SB}^2 - (|\hat{\bm{s}}_{y}| - |\hat{\bm{l}}_{\mathrm{b},y}|)^2 = 2(1-\cos\theta_\mathrm{sb}) - (\cos \theta_\mathrm{sd} - \cos \theta_\mathrm{db})^2,
\end{align}
and thus
\begin{align}
    \cos \Delta \Omega &= \frac{\sin^2\theta_\mathrm{db} + \sin^2\theta_\mathrm{sd} + (\cos \theta_\mathrm{sd} - \cos \theta_\mathrm{db})^2 - 2(1-\cos\theta_\mathrm{sb})}{2\sin\theta_\mathrm{db}\sin\theta_\mathrm{sb}},
\end{align}
which simplifies to Eq.~\eqref{eq:rel_longitude}.

\bibliography{references}{}
\bibliographystyle{aasjournal}

\end{document}